\newtheorem{definition}{Definition}
\newtheorem{theorem}{Theorem}
\newcommand{\bt}{\begin{theorem}}
\newcommand{\et}{\end{theorem}}
\newcommand{\bd}{\begin{definition}}
\newcommand{\ed}{\end{definition}}
\newcommand{\be}{\begin{equation}}
\newcommand{\ee}{\end{equation}}
\newcommand{\bear}{\begin{eqnarray}}
\newcommand{\eear}{\end{eqnarray}}
\newcommand{\baar}{\begin{array}}
\newcommand{\eaar}{\end{array}}
\newcommand{\nn}{\nonumber}
\newcommand{\pr}{\partial}
\newcommand{\G}{\Gamma}

\documentclass[11pt]{article}

\usepackage{latexsym}
\usepackage{amsfonts}
\usepackage[dvips]{graphicx}

\begin{document}
\renewcommand{\thesection}{\Roman{section}.}
\renewcommand{\thesubsection}{\Alph{subsection}.}
\thispagestyle{empty} 
\setlength{\topmargin}{-20mm}
\setlength{\textheight}{250mm}
{\begin{center}
\large{\bf Janet-Riquier Theory and the Riemann-Lanczos Problems in 2 and 
           3 Dimensions}
\end{center}}
\vspace{0.5cm}\noindent
P Dolan$^{\: a)}$\\
Mathematics Department, Imperial College, 180 Queen's Gate,\\
London SW7 2BZ
\newline
\vspace{0.5cm}
\newline
A Gerber$^{\: b)}$\\
Centre for Techno-Mathematics and Scientific Computing Laboratory,\\
University of Westminster, Watford Road, Harrow HA1 3TP
\vspace{0.5cm}\indent
\begin{center} 
\large{Abstract} 
\end{center}

The Riemann-Lanczos problem for 4-dimensional manifolds was discussed by 
Bampi and Caviglia. Using exterior differential systems they showed that it 
was not an involutory differential system until a suitable prolongation was 
made. Here, we introduce the alternative Janet-Riquier theory and use it to 
consider the Riemann-Lanczos problem in 2 and 3 dimensions.
We find that in 2 dimensions, the Riemann-Lanczos problem is a differential 
system in involution. It depends on one arbitrary function of 
2 independent variables when no differential gauge condition is imposed but
on 2 arbitrary functions of one independent variable when the differential 
gauge condition {\bf is} imposed. For each of the two possible signatures we 
give the general solution in both instances to show that the occurrence of
characteristic coordinates need not affect the result.
In 3 dimensions, the Riemann-Lanczos problem is not in involution as a 
so-called ``internal'' identity occurs. This does not prevent the existence of
singular solutions. A prolongation of this problem, where an integrability 
condition is added, leads to an involutory prolonged system and thereby 
generates {\it non-singular} solutions of the prolonged Riemann-Lanczos 
problem. We give a singular solution for the unprolonged Riemann-Lanczos 
problem for the 3-dimensional reduced G\"{o}del spacetime.
\vspace{0.2cm}
\footnoterule
\vspace{0.2cm}\noindent
$^{a)}$ Electronic mail: pdolan@inctech.com\\
$^{b)}$ Electronic mail: a\_gerber01@hotmail.com
\newpage
\indent
\setlength{\arraycolsep}{0.5mm}
\setlength{\topmargin}{-20mm}
\section{Introduction}
The problem of generating the spacetime Weyl conformal curvature tensor 
$C_{abcd}$ from a tensor potential is called {\it the Weyl-Lanczos problem} 
and the analogous problem for the Riemann curvature tensor {\it the 
Riemann-Lanczos problem}.

The work in this paper is based on the papers \cite{Bam1,Bam2} and on 
\cite{DoGe1} for the Riemann-Lanczos problem as an exterior differential 
system. But here, for simplicity, we mainly look at the Riemann-Lanczos 
problem in 2 and 3 dimensions as an exterior differential system and then use 
Janet-Riquier theory to verify our results. As already explained in many 
papers such as \cite{Bam1,Bam2,DoGe1} we can express the Riemann tensor by 
means of a third-order tensor potential called the Lanczos tensor with 
components $L_{abc}$ which are subject to the following symmetries 
\be
L_{[abc]}=0 \label{1cyclic}
\ee
and
\be
L_{abc}=L_{[ab]c} \: . \label{1syms}
\ee
If (\ref{1cyclic}) and (\ref{1syms}) are imposed, we obtain 2 independent 
components in 2 dimensions, 8 independent in 3 dimensions and 20 in 4 
dimensions. The Riemann-Lanczos equations were first published by Udeschini
Brinis \cite{Ude1} as
\be
R_{abcd} = L_{abc;d}-L_{abd;c}+L_{cda;b}-L_{cdb;a} \: , \label{1Rieml}
\ee
where ``;'' denotes covariant differentiation.
However, as explained in \cite{Bam1,Bam2}, they did not form a system in 
involution but had to be prolongated once in order to be in involution. 
In what follows we include the differential gauge conditions
\be
{L_{ab}}^{s}_{\: ;s}=0 \label{1diffg}
\ee
{\bf but not} the trace-free gauge conditions 
\be
{L_{a}}^{s}_{\: s}=0 \label{1trace} \: .
\ee
If equations (\ref{1trace}) were to be imposed, we would obtain the equation
$$ R = 4 {L^{nk}}_{k;n} = -4 {L^{nk}}_{n;k} = 0 $$
for the Ricci scalar $R$ which leads to an inconsistent theory. 
\section{Janet-Riquier Theory}
In the original Janet-Riquier theory of systems of partial differential 
equations (PDEs), there was an algorithm created to explain how a given
system of PDEs could be brought into passive form. Passivity is the 
{\it absence} of integrability conditions and orthonomicity is a form 
of ordering the partial derivatives of a system. The passive orthonomic system
of PDEs was the predecessor of what is now called a formally integrable system
or involutory system of PDEs.

For a system of partial differential equations it is always an important 
question whether it possesses a general solution or whether further 
conditions have to be imposed so that the system is formally 
integrable or in involution. A theory created by Riquier \cite{Riqu}
and developed by Janet \cite{Jan1,Jan2} and Thomas \cite{Tom1} now called 
Janet-Riquier theory helps to decide such questions. Applications of the work 
of Riquier to general relativity have previously been made in papers by
Pereira \cite{Per1,Per2,Per3}\footnotemark[1].
A good account of this theory as in the form as created in \cite{Riqu,Jan1,
Tom1} is given in Russian by S.P. Finikov \cite{Fini}.

The original approach by Janet, Riquier and Thomas however did not lead to 
intrinsic results. To achieve this Spencer \cite{Spen} and then Goldschmidt 
\cite{Gold} introduced a new coordinate independent approach based on 
homological algebra. An account of this theory can be found in \cite{Pom2}.
The earlier approach by Thomas \cite{Tom1} was thorough but based on elaborate
systems of inequalities for monomials corresponding to a system of PDEs. 

There are many algorithms which have been implemented into algebraic 
computing. Some of them are concerned with the choice of particular rankings 
such as the REDUCE package DINV by Gerdt \cite{Gert}. Reid \cite{Rei2} 
developed a computer package based on MAPLE which brings a system into {\it 
solved form} which is a modification of an orthonomic system \cite{Rei1}. 
Seiler \cite{Sei1,Sei2} uses the theory of the involutive symbol,
which is a modernised version of the original Janet-Riquier theory, to 
determine whether systems are in involution.

Next, we shall explain a few terms which are important in the original as well 
as in modernised versions of Janet-Riquier theory. For a system of partial 
differential equations (=system of PDEs) of order $q$, which we shall denote by
$\mathcal{R}_{q}$ from now on, we use $x^{1},\cdots ,x^{n}$ for the 
independent variables, where we shall use brackets to indicate powers of any 
$x^{i}$ such as in ${(x^{i})}^{n}$, and $u^{\alpha}$ for the $m$ 
dependent variables. Their derivatives on the jet bundle of 
order $q$, denoted by $\mathcal{J}^{q}(\mathbb{R}^{n},\mathbb{R}^{m})$, are 
denoted by $u^{\alpha}_{,J}$, where $J$ is a multi-index. 
We associate a {\bf monomial} $x^{J}=(x^{1})^{j_{1}}\cdot ... 
\cdot (x^{n})^{j_{n}}$ with each partial derivative $u^{\alpha}_{,J}$, where 
$J = (j_{1},\cdots , j_{n})$ is a {\bf multi-index} and 
$\| J \| =j_{1} + \cdots +j_{n}$ is the order of the derivative. 
This means that to each set of partial derivatives 
of each dependent variable $u^{\alpha}$ corresponds a unique set of monomials.

Then, we {\bf order the partial derivatives} of a system of PDEs
in a systematic way. Very often when $x^{1},x^{2}, \cdots , x^{n}$ are
our independent variables, we shall use an inverse lexicographic ordering 
based on $x^{n} \succ \cdots \succ x^{2} \succ x^{1}$ on their partial 
derivatives so that $u_{,n} \succ \cdots \succ u_{,2} \succ u_{,1}$ and so 
on. Generally, we introduce a {\bf ranking} amongst {\it all} partial 
derivatives 
\bd{Ranking of Derivatives}\\
A ranking of derivatives is a total ordering $\mathcal{R}_{\succ}$ of all the 
partial derivatives $u^{\alpha}_{,J}$ (for $m,n$ fixed) satisfying the two 
conditions (where $J$ and $K$ are multi-indices)\\
i) If $\|J\| > \| K \|$, then $u^{\alpha}_{,J} \succ u^{\alpha}_{,K}$.\\
ii)If $u^{\alpha}_{,J^{1}} \succ u^{\alpha}_{,J^{2}}$, then 
      $(u^{\alpha}_{,J^{1}})_{,K} \succ (u^{\alpha}_{,J^{2}})_{,K}$ for any
multi-index $K$.
\ed
A special subclass of these rankings are called {\bf orderly rankings} by 
which we mean that these are rankings such that i) holds for different 
indices $\alpha$ referring to different unknowns $u^{\alpha}$. A more detailed 
account on the problem of finding a suitable ranking especially for 
non-linear systems of equations is given by Rust \cite{Rus1}. Every ranking 
defined on the partial derivatives induces a ranking amongst the monomials 
which correspond to the given partial derivatives.

Once a ranking for a system of equations $\mathcal{R}_{q}$ is determined, the 
system can be brought into a more organised form called {\bf orthonomic form}. 
This is achieved by determining the partial derivative highest in the ranking 
in each equation of $\mathcal{R}_{q}$ and calling it the {\bf leading} 
derivative of the equation. Once this equation is solved for its leading 
derivative which then becomes the only term on its LHS, we call it a 
{\bf principal} derivative. All other partial derivatives of that order which 
are not in the set of principal derivatives are called {\bf parametric 
derivatives}. Based on this we can define an
\bd{Orthonomic System} \label{Deffi}\\
A system of partial differential equations $\mathcal{R}_{q}$ of order 
q is orthonomic with respect to a given ranking $\mathcal{R}_{\succ}$ if\\
i) all the PDEs are solved with respect to their leading derivatives;\\
ii) no two leading derivatives are the same;\\
iii) and no parametric derivative in any equation of $\mathcal{R}_{q}$ can 
be a principal derivative in another equation of $\mathcal{R}_{q}$ or even a 
partial derivative of any order of a principal derivative.
\ed
We base our calculations on definition (\ref{Deffi}) as our own 
standard. For orthonomic systems the only term on each LHS is the principal 
derivative so that all derivatives of order $q$ on the RHSs of the equations 
are parametric derivatives. By differentiating each such equation we can also 
get the derivatives of the principal derivatives in such a form that we can 
define {\bf multiplicative variables} for each equation. Those independent 
variables by means of which we can differentiate the principal derivative of 
an equation without re-introducing a derivative already produced by means of 
differentiating another principal derivative are called 
{\bf multiplicative variables} for 
the equation. In the same way, once we associate a monomial with each 
principal derivative, we can define what the multiplicative variables are for 
that monomial. We say that a variable $x^{n}$ is multiplicative for the 
monomials which are of maximal degree in $x^{n}$. Further, $x^{i}$ with 
$i < n$ is multiplicative for $x^{J}=(x^{1})^{j_{1}} \cdots (x^{n})^{j_{n}}$ 
if amongst all monomials of the form $(x^{1})^{k_{1}} \cdots (x^{i})^{k_{i}} 
(x^{i+1})^{j_{i+1}} \cdots (x^{n})^{j_{n}}$ the monomial $x^{J}$is 
{\bf such that} $j_{i}=\mbox{max}_{k} k_{i}$. This definition given in 
\cite{Pom2} is based on Janet's original definition \cite{Jan1}.

We call a set of monomials {\it complete} if any multiple can be obtained 
using multiplicative variables only. We can also define completeness using
partial derivatives. Given an orthonomic system, we denote the set of all 
leading derivatives and their derivatives with respect to multiplicative 
variables only of the same dependent variable $u^{\alpha}$ by 
$\mathcal{L}^{\alpha}_{\succ}$.
The closure $\bar{\mathcal{L}}^{\alpha}_{\succ}$ then includes all derivatives 
of the derivatives of $u^{\alpha}$. 
We call a system of partial differential equations of order $q$, 
$\mathcal{R}_{q}$, a {\bf complete system} when $\mathcal{L}^{\alpha}_{\succ}
=\bar{\mathcal{L}}^{\alpha}_{\succ}$ $\mbox{for all} \: u^{\alpha} \: 
, \: \alpha = 1, \cdots , n \: .$
Once our system is complete, we can also ask whether it is {\it passive}.
A system of PDEs having a complete set of monomials is called {\bf passive} 
if any computation of a principal derivative is equivalent to any computation
obtained using multiplicative variables only. Otherwise, the additional 
equations have to be added as integrability conditions.
If a complete system of PDEs is given in orthonomic form, we speak of a
\bd{Passive Orthonomic System}\\
An orthonomic system of equations $\mathcal{R}_{q}$ is passive with 
respect to a ranking $\mathcal{R}_{\succ}$, if the sets of all leading 
derivatives $\mathcal{L}^{\alpha}_{\succ}$ with respect to the chosen 
ranking $\mathcal{R}_{\succ}$ are complete and {\bf no integrability 
conditions} occur.
\ed
Riquier \cite{Riqu} formulated an important theorem for the local 
existence of analytic solutions based on this, namely, 
\bt{Riquier}\\
Given a passive system with respect to an orderly ranking 
$\mathcal{R}_{\succ}$, then its formal power series solution converges.
\et
The main difficulty which arises in Janet-Riquier theory is that so many
results are not intrinsic and depend on the choice of a ranking.
In a more recent version this problem is re-examined by introducing
special coordinates called {\bf $\delta$-regular coordinates} and by using the
theory of the {\bf involutive symbol} of a system $\mathcal{R}_{q}$ which we 
shall introduce next. But before that, we shall discuss a Janet example in 
detail in order to illustrate the theory exhibited so far.

A classical example of Janet \cite{Jan2,Sei1,Pom2} consists of two second-order
equations with 3 independent variables $x^{1},x^{2},x^{3}$ and one dependent 
variable ${\it u}$ so that $n=3$, $m=1$ and $q=2$. The two equations are 
\bear
u_{,33}-x^{2}u_{,11} & = & 0\: , \nn\\
u_{,22} & = & 0 \: . \label{Janne}
\eear
If we impose the coordinate ranking $x^{3} \succ x^{2} \succ x^{1}$, we 
obtain the ranking $u_{,33} \succ u_{,23} \succ u_{,22} \succ u_{,13} 
\succ u_{,12} \succ u_{,11}$ amongst the second-order partial derivatives. 
The leading derivatives then are $u_{,33}$ in the first equation and $u_{,22}$
in the second equation so that the system in orthonomic form is given by
\bear
u_{,33} & = & x^{2} u_{,11} \qquad  x^{1} \: x^{2} \: x^{3} \nn\\
u_{,22} & = & 0 \qquad \qquad x^{1} x^{2} \bullet \: . \label{Janete}
\eear 
The principal derivatives are $u_{,33}$ and $u_{,22}$ which have the 
corresponding set of monomials $\{(x^{3})^{2},(x^{2})^{2} \}$ and 
the parametric derivatives are $u_{,11}, u_{,12}, u_{,13},$\\ 
$u_{,23}$. 
In the first equation all variables $x^{1},x^{2},x^{3}$ are multiplicative 
variables but only $x^{1}$ and $x^{2}$ in the second equation whereas the
variable $x^{3}$ is not a multiplicative variable in the second equation 
(which is indicated by a dot). This is because, by means of differentiating 
the first equation twice with respect to $x^{2}$, $u_{,2233}$ and higher 
derivatives can already be created\footnotemark[1].
For the Janet example (\ref{Janne}), its completion is given in 
\cite{Pom2},\cite{Sei1}. When choosing the above ranking based on $x^{3} 
\succ x^{2} \succ x^{1}$, we obtain for $\mathcal{L}_{\succ}$ and for 
$\bar{\mathcal{L}}_{\succ}$
$$\mathcal{L}_{\succ}= \{ u_{,33};u_{,22};u_{,133};u_{,233};u_{,333};u_{,122};
u_{,222} \}$$
$$\bar{\mathcal{L}}_{\succ}= \{ u_{,33};u_{,22};u_{,133};u_{,233};u_{,333};
u_{,122};u_{,222};u_{,223} \} \: .$$
Therefore, $u_{,223}$ cannot be produced using multiplicative variables only
- a fact reflected by the impossibility of producing the monomial $(x^{2})^{2}
\cdot x^{3}$ using the set $\{ (x^{3})^{2},(x^{2})^{2} \}$. We have to add 
$u_{,223}$ so that the new completed system is
\[ \baar{lll}
u_{,223} = & 0 \qquad & x^{1} \: x^{2} \: \bullet \nn\\
 & & \nn\\
u_{,33} - x^{2} u_{,11} = & 0 \qquad & x^{1} \: x^{2} \: x^{3} \nn\\
 & & \nn\\
u_{,22} = & 0 \qquad & x^{1} \: x^{2} \: \bullet \: .
\eaar \]
But this system is not passive because, using multiplicative variables only,
we can form $u_{,2233}-x^{2}u_{,1122}-2u_{,112}=0$ which leads to the 
integrability condition $u_{,112}=0$. We refer the reader to \cite{Pom2}
for a detailed discussion of the completion process and only state the final 
prolonged complete system as
\[ \baar{lll}
u_{,11113} = & 0 \qquad & x^{1} \: \bullet \: \bullet \nn\\
 & & \nn\\
u_{1123} = & 0 \qquad & x^{1} \: \bullet \: \bullet \nn\\
 & & \nn\\
u_{,1111} = & 0 \qquad & x^{1} \: \bullet \: \bullet \nn\\
 & & \nn\\
u_{,223} = & 0 \qquad & x^{1} \: x^{2} \: \bullet \nn\\
 & & \nn\\
u_{,112} = & 0 \qquad & x^{1} \: \bullet \: \bullet \nn\\
 & & \nn\\
u_{,33} -x^{2} u_{,11} = & 0 \qquad & x^{1} \: x^{2} \: x^{3} \nn\\
 & & \nn\\
u_{,22} = & 0 \qquad & x^{1} \: x^{2} \: \bullet \: ,
\eaar \]
where {\it prolongation} of a system of equations $\mathcal{R}_{q}$ simply
means adding all derivatives of order $q+1$ of the equations in 
$\mathcal{R}_{q}$ leading to the prolonged system of equations 
$\mathcal{R}_{q+1}$, a process which can be carried out several times if 
necessary.
\subsection{Involutive Symbol and Formal Integrability}
In modern versions following Spencer \cite{Spen} one wishes to replace the 
coordinate-dependent theory of complete passive orthonomic systems of 
equations $\mathcal{R}_{q}$ by means of quantities which will no longer 
depend on a choice of coordinates. A theoretical approach based on map 
diagrams from homological algebra \cite{Spen,Gold} also explained in 
\cite{Pom2} does fulfil these requirements. 

For practical purposes though we shall rely on the theory of the {\bf 
involutive symbol} and on {\bf $\delta$-regular coordinates}. The {\bf symbol}
of a system of PDEs involves only the highest-order partial derivatives of 
each equation in $\mathcal{R}_{q}$ which simplifies calculations especially of
large systems significantly. We denote the quantities corresponding to each 
partial derivative $u^{\alpha}_{,J}$ by $V^{\alpha}_{J}$ and define the symbol
$\mathcal{M}_{q}$ as
\bd{Symbol of $\mathcal{R}_{q}$}\\
A system of partial differential equations $\mathcal{R}_{q}$ of order q
locally described by $p$ equations in solved form as $\Phi^{\tau}(x^{i},
u^{\alpha},u^{\alpha}_{,J})=0$ for $\tau =1,...,p,$ has a solution space 
$\mathcal{M}_{q}$ for the unknowns $V^{\alpha}_{J}$ with 
$\alpha =1,...,m$, $\| J \|=q$:
\bear 
\mathcal{M}_{q}: {\sum_{\alpha,\| J \|=q}}\left( \frac{\pr \Phi^{\tau}}
{\pr u^{\alpha}_{,J}}\right) V^{\alpha}_{J} & = & 0 \: , \label{2Symm}
\eear
where we {\it formally differentiate} with respect to the $u^{\alpha}_{,J}$.
$\mathcal{M}_{q}$ is called the {\bf symbol of $\mathcal{R}_{q}$}.
\ed
For simplicity the matrix rather than the map is usually regarded as the symbol
of $\mathcal{R}_{q}$. We associate with each symbol equation its 
multiplicative variables which are the same as the multiplicative variables 
its corresponding equation in $\mathcal{R}_{q}$ adopts. Once each symbol 
equation has its multiplicative variables allocated, we determine the 
{\bf class of an equation} in $\mathcal{M}_{q}$ by counting the number of 
multiplicative variables it adopts - a number denoted by $k$ such that 
$0 \leq k \leq n$. We carry this out with all equations occurring in 
$\mathcal{M}_{q}$, counting how many equations of each class there are. So we 
define
$$ \beta^{(k)}_{q} := \mbox{number of equations of class} \: k \: \mbox{in} \:
\mathcal{M}_{q} \: (\mbox{or}\mathcal{R}_{q}) \: .$$
The definition of the {\bf Cartan characters} \footnotemark[2]
$\alpha^{(k)}_{q}$is based on the $\beta^{(k)}_{q}$ and given by
\be
\alpha^{(k)}_{q} := m\cdot{{n+q-k-1}\choose{q-1}}-\beta^{(k)}_{q}
\label{2Alfa} \: ,
\ee
where $m\cdot \sum^{n}_{k=1}{{n+q-k-1}\choose{q-1}}$
is the total number of partial derivatives of order $q$ that a 
system $\mathcal{R}_{q}$ will have. The Cartan character $\alpha^{(k)}_{q}$ 
represents the number of remaining independent partial derivatives 
of order $q$ in the subsystem of class $k$ after the removal of the number of 
the principal derivatives of class $k$ in that subsystem. In $\delta$-regular
coordinates $\alpha^{(k)}_{q}$ equals the number of parametric derivatives of
class $k$ and order $q$.

Because the notion ``class of an equation'' is {\bf ranking-dependent},
we must ensure that we obtain intrinsic results. This can be achieved 
when we use a ranking for which we obtain the maximal possible value 
for $\beta^{(n)}_{q}$ and then for $\beta^{(n-1)}_{q}+\beta^{(n)}_{q}$, 
and so on until we obtain the maximal possible value for 
$\sum^{n}_{k=1}\beta^{(k)}_{q}$. This is equivalent to obtaining the 
minimal possible value for $\alpha^{(n)}_{q}$, then for 
$\alpha^{(n-1)}_{q} + \alpha^{(n)}_{q}$ and so on until the minimal 
possible value for $\sum^{n}_{k=1}\alpha^{(k)}_{q}$ is achieved. 
In an arbitrarily given coordinate system, a prolongation to higher order is
sometimes necessary to obtain intrinsic results. This can be avoided by
performing a coordinate transformation, where a linear coordinate 
transformation is sufficient \cite{Pom2}, and by checking again 
for minimal and maximal values respectively of the $\alpha^{(k)}_{q}$ and the 
$\beta^{(k)}_{q}$ in the above sense. Once this is fulfilled, we are using a 
system of {\bf $\delta$-regular coordinates}.

It is important to determine whether a given system of equations 
$\mathcal{R}_{q}$ possesses identities or not. When no identities are 
present, the symbol is said to be involutive which can be equivalently 
expressed as \cite{Pom2}
\bt{Involutive Symbol}\\
In a $\delta$-regular coordinate system the following conditions are
equivalent:\\
i) The symbol $\mathcal{M}_{q}$ is involutive $\:$;\\
ii) dim $(\mathcal{M}_{q+1}) = \sum^{n}_{k=1} k \cdot \alpha^{(k)}_{q} \: ;$\\
iii) for the rank $r$ of $\mathcal{M}_{q+1}$ it is $r(\mathcal{M}_{q+1}) 
= \sum^{n}_{k=1} k \cdot \beta^{(k)}_{q} \: ;$\\
iv) prolongation with respect to {\bf non-multiplicative variables} does not 
lead to any new equations.
\et
But a system of equations $\mathcal{R}_{q}$ which has an involutive symbol 
$\mathcal{M}_{q}$ can still admit integrability conditions. They can be 
revealed by means of projecting our prolonged system $\mathcal{R}_{q+1}$,  
which is obtained by differentiating $\mathcal{R}_{q}$ with 
respect to all its $n$ independent variables, back onto $\mathcal{R}_{q}$. In 
general, we shall denote first projections onto lower-order systems by 
$\mathcal{R}^{(1)}_{q+r} = \pi^{(q+r+1)}_{q+r}(\mathcal{R}_{q+r+1})$. 
If $\mathcal{R}^{(1)}_{q}$ is not identical to $\mathcal{R}_{q}$, then
integrability conditions occur, but if they are identical we can characterise 
a formally integrable system $\mathcal{R}_{q}$ as \cite{Sei1,Pom2}
\bd{Formal Integrability}\\
A system of partial differential equations $\mathcal{R}_{q}$ is formally 
integrable means that $\mathcal{R}^{(1)}_{q+r}=\mathcal{R}_{q+r}$ for 
all $r \geq 0 \: .$
\ed
A special situation occurs when $\mathcal{M}_{q}$ is involutive {\it and} 
$\mathcal{R}_{q}$ is formally integrable \cite{Sei1}. Then,
we obtain ideal systems of equations which are called
\bd{Involutive Systems of Equations}\\
A system of equations $\mathcal{R}_{q}$ is involutive if and only if
its symbol $\mathcal{M}_{q}$ is involutive and $\mathcal{R}_{q}$ is 
formally integrable.
\ed
Note that if we are dealing with a vector bundle, it is always valid 
$\mbox{dim}(\mathcal{R}^{(1)}_{q})=\mbox{dim} (\mathcal{R}_{q+1}) -\mbox{dim} 
(\mathcal{M}_{q+1})$. This means that when we know the dimension of the 
prolonged system of equations $\mathcal{R}_{q+1}$ and the dimension of 
its prolonged symbol $\mathcal{M}_{q+1}$, we automatically know the 
dimension of the system of equations projected back onto order $q$ called 
$\mathcal{R}^{(1)}_{q}$. This can be a great help in deciding whether a system
has integrability conditions.

The subsequent theorems, which can be found in \cite{Pom2,Sei1}, are very 
useful for practical applications as only a limited number of steps need to 
be carried out to check for involutivity. First we look at
\bt
Let $\mathcal{M}_{q}$ be an involutive symbol of a system of equations
$\mathcal{R}_{q}$, then\\
i) the symbol $\mathcal{M}_{q+1}$ is also involutive, and\\
ii) ${(\mathcal{R}^{(1)}_{q})}_{+1}=\mathcal{R}^{(1)}_{q+1} \: .$ 
\et
Here i) states that, for any involutive symbol $\mathcal{M}_{q}$, a 
prolongation is trivial and the prolonged symbol is again involutive 
and (ii) says that in this case projections and 
prolongations commute. A proposition following from this is 
\bt{Proposition}\\
Given a system of differential equations $\mathcal{R}_{q}$, where the number
of equations equals the number of dependent variables $m$, then, 
$\mathcal{R}_{q}$ has either identities or integrability 
conditions which implies that $\beta^{(n-1)}_{q} > 0.$ 
\et
Based on the previous definitions, we would have to perform infinitely 
many prolongations in order to decide whether a given system $\mathcal{R}_{q}$ 
is involutive or not. Fortunately this is not necessary and we can use the 
important
\bt{Criterion for Involution}\\
$\mathcal{R}_{q}$ is a system in involution means that its symbol 
$\mathcal{M}_{q}$ is involutive \underline{and} $\mathcal{R}^{(1)}_{q}
=\mathcal{R}_{q}$.
\et

We shall now present the modern version of Janet-Riquier theory as given above
using the Janet example (\ref{Janne}). Its symbol $\mathcal{M}_{2}$ is given by
\[ \baar{ll|r}
V_{33} = x^{2}V_{11} \qquad & x^{1} \: x^{2} \: x^{3} \quad & \quad 3 \nn\\
V_{22} = 0 \qquad & x^{1} \: x^{2} \: \bullet \quad & \quad \: 2 ,
\eaar \]
where the integers on the RHS of each equation indicate the class of this
equation. This leads to $\beta^{(1)}_{2}=0, \beta^{(2)}_{2}= 
\beta^{(3)}_{2}=1$ and so $\sum^{3}_{k=1}k\cdot\beta^{(k)}_{2}=5$ which 
gives us the {\bf total number of multiplicative variables} for 
$\mathcal{M}_{2}$ to be 5. In order for $\mathcal{M}_{2}$ to be involutive, we
have to verify that $r(\mathcal{M}_{3})$ is equal to 5. Firstly, we determine 
the new prolonged system of equations $\mathcal{R}_{3}$ which consists of 
the 8 equations 
\[ \baar{ll}
\fbox{1} \qquad & 0 = u_{,33}-x^{2}u_{,11} \nn\\
\fbox{2} \qquad & 0 = u_{,22} \nn\\
\fbox{3} \qquad & 0 = u_{,133} - x^{2} u_{,111} \nn\\
\fbox{4} \qquad & 0 = u_{,233} - x^{2} u_{,112} - u_{,11} \nn\\
\fbox{5} \qquad & 0 = u_{,333} - x^{2} u_{,113} \nn\\
\fbox{6} \qquad & 0 = u_{,122} \nn\\
\fbox{7} \qquad & 0 = u_{,222} \nn\\
\fbox{8} \qquad & 0 = u_{,223} \: . \label{2Jan3}
\eaar \]
Next, we determine $\mbox{dim}(\mathcal{R}_{3})$, where $\mbox{dim}
(\mathcal{R}_{q})$ is defined as the total number of dependent 
variables and all their partial derivatives up to order $q$ {\bf minus} the 
number of independent equations which constitute $\mathcal{R}_{q}$. In this 
way, we can easily see that $\mbox{dim}(\mathcal{R}_{3})=20-8 =12$, where $8$ 
is the total number of independent equations in $\mathcal{R}_{3}$. The number 
20 stands for $20=1+3+6+10$ for one unknown $u$, 3 first-order partial 
derivatives $u_{,1},u_{,2},u_{,3}$, 6 second-order partial derivatives 
$u_{,11},u_{,12},u_{,13},u_{,22},u_{,23},u_{,33}$ and 10 third-order partial 
derivatives $u_{,111},u_{,112},u_{,113},u_{,122},u_{,123},u_{,133},u_{222},
u_{223},u_{,233},u_{,333}$. Next, we determine $\mathcal{M}_{3}$ which
is given in orthonomic form as
\[ \baar{ll|r}
V_{333} = x^{2}V_{113} \qquad & \: x^{1} \: x^{2} \: x^{3} \quad & \quad 3 
\nn\\
V_{233} = x^{2}V_{112} \qquad & \: x^{1} \: x^{2} \: \bullet \quad & \quad 2 
\nn\\
V_{223} = 0 \qquad & \: x^{1} \: x^{2} \: \bullet \quad & \quad 2 \nn\\
V_{222} = 0 \qquad & \: x^{1} \: x^{2} \: \bullet \quad & \quad 2 \nn\\
V_{133} = x^{2}V_{111} \qquad & \: x^{1} \: \bullet \: \bullet \quad & \quad 1
\nn\\
V_{122} = 0 \qquad & \: x^{1} \: \bullet \: \bullet \: \quad & \quad 1 ,
\eaar \]
where we can see that there are 2 equations of class 1, 3 equations of 
class 2 and 1 equation of class 3 resulting in $\beta^{(1)}_{3}=2, 
\beta^{(2)}_{3}=3, \beta^{(3)}_{3}=1$. This leads to 
$\sum^{3}_{k=1} k\cdot \beta^{(k)}_{3}=1 \cdot 2 + 2 \cdot 3 + 3 \cdot 1 =11$ 
being the total number of multiplicative variables which $\mathcal{M}_{3}$ 
adopts. We also find that $r(\mathcal{M}_{3})=6$ which is not equal the 
total number of multiplicative variables $\mathcal{M}_{2}$ adopts, which is 5,
and therefore $\mathcal{M}_{3}$ is not involutive. We need to prolongate
$\mathcal{R}_{3}$ further to $\mathcal{R}_{4}$ and we refer the reader to 
\cite{Ger1} for further details subsequent calculations. We only state here 
that during this process the two integrability conditions
\bear
u_{,112} & = & 0 \nn\\
u_{,1111} & = & 0 \label{Iintt}
\eear
occur. Finally, we obtain the result that the prolonged fifth-order system, 
which is augmented by (\ref{Iintt}), $\mathcal{R}^{(2)}_{5}$, forms a system 
in involution and the Janet algorithm terminates. 
\section{The Riemann-Lanczos Problem in 2 Dimensions}
We shall apply the theory of EDS and Janet-Riquier theory in order to
examine the Riemann-Lanczos problems in 2 and in 3 dimensions. For a review
of the theory of EDS see \cite{Cart,5man,Sleb,Yang,Ger1,DoGe1}. 

We show that the Riemann-Lanczos problem is in involution with respect to its 
two independent variables for both possible choices of signature and show that
the occurrence of characteristic coordinates need not affect the result. We 
shall look at the problem first as an exterior differential system and then as 
a system of partial differential equations (systems of PDEs) applying 
Janet-Riquier theory.
\subsection{The Riemann-Lanczos Problem in 2 Dimensions as an EDS}
Firstly, we consider the Riemann-Lanczos problem in a 2-dimensional spacetime. 
We know that in 2 spacetime dimensions, there is only one independent 
component of the Riemann tensor and therefore only one independent 
Riemann-Lanczos equation. We now choose $(x^{1},x^{2},L_{121},
L_{122},P_{1211},P_{1212},$\\
$P_{1221},P_{1222})$ as local coordinates on 
the jet bundle $\mathcal{J}^{1}(\mathbb{R}^{2},\mathbb{R}^{2})$ so that 
it is 8-dimensional. We leave the
metric tensor components completely arbitrary so that the following 
results will hold for any 2-dimensional spacetime. 
The only independent Riemann-Lanczos equation is given in solved form as
\bear
f^{(R)}_{1212} & = & R_{1212} -2P_{1212} + 2P_{1221} 
         +2L_{121}(\Gamma^{1}_{12}-\Gamma^{2}_{22})
         -2L_{122}(\Gamma^{1}_{11}+\Gamma^{2}_{12}) \nn\\
& = & 0 \: .
\label{5Rie2}
\eear
We define $\alpha_{1212e}$ in the 2-dimensional case to be
\be
\alpha_{1212e} = R_{1212,e}
 -2L_{121}(\G^{1}_{12,e}-\G^{2}_{22,e})
 -2L_{122}(\G^{1}_{11,e}+\G^{2}_{12,e})
\ee
where $e=1,2$. The Pfaffian system derived from (\ref{5Rie2}) is locally 
given by
\bear
\theta^{1} & = & \alpha_{12121}d x^{1}+\alpha_{12122}d x^{2}- 2d P_{1212} 
+2 d P_{1221}+2(\Gamma^{1}_{12}-\Gamma^{2}_{22})d L_{121} \nn\\
& & -2(\Gamma^{1}_{11}+\Gamma^{2}_{12})d L_{122} \: , \nn\\
\theta^{2} & = & d L_{121}- P_{1211}d x^{1} -P_{1212} d x^{2} \: , \nn\\
\theta^{3} & = & d L_{122}- P_{1221}d x^{1} -P_{1222} d x^{2} \: , \nn\\
d \theta^{2} & = & d x^{1}\wedge \ P_{1211} + d x^{2} \wedge P_{1212} \: ,\nn\\
d \theta^{3} & = & d x^{1}\wedge \ P_{1211} + d x^{2} \wedge P_{1212} \: ,
\label{5REDS2}
\eear
where $\theta^{1}$ is the exterior derivative of $f^{(R)}_{abcd}$ and
$\theta^{2}, \theta^{3}$ are the two contact conditions and 
$d \theta^{2},d \theta^{3}$ their exterior derivatives. 
The zeroth character is $s_{0}=3$, since we only have to 
count the number of 1-forms in (\ref{5REDS2}). Omitting all the terms 
involving $d x^{1}, d x^{2}$ in the 1-forms in \ref{5REDS2}, the same number 
is obtained so that $s_{0}'=3$ leading to our first result $s_{0}=s_{0}'=3$. 
In order to obtain $s_{1}$, we must form the first polar element 
$H((E^{1})_{x})$ of a 1-dimensional integral element $(E^{1})_{x}$ formed by 
the Vessiot vector field $U$ 
\be
U= U^{e}\frac{\pr}{\pr x^{e}}+ P_{121e}U^{e}\frac{\pr}{\pr L_{121}}
+P_{122e}U^{e}\frac{\pr}{\pr L_{122}}+U_{121e}\frac{\pr}{\pr P_{121e}}
+U_{122e}\frac{\pr}{\pr P_{122e}} \: ,
\ee
where
\be
U_{1212} = U_{1221} + \delta_{1}U^{1} + \delta_{2}U^{2}\\
\ee
because
\bear
\delta_{1} & = & \frac{1}{2}\alpha_{12121}
+ (\G^{1}_{12}-\G^{2}_{22})P_{1211}
- (\G^{1}_{11}+\G^{2}_{12})P_{1221} \: , \nn\\
\delta_{2} & = & \frac{1}{2}\alpha_{12122}
+ (\G^{1}_{12}-\G^{2}_{22})P_{1212}
- (\G^{1}_{11}+\G^{2}_{12})P_{1222} \: .
\eear
This means that for the Vessiot vector field $U$ the components 
$\: U^{1},U^{2},U_{1211},$ \\
$U_{1221},U_{1222}$ can be chosen arbitrarily and so we can form 
$H((E^{1})_{x})$. For the coefficient matrix of the 
second polar system $H((E^{2})_{x})$, we obtain
\[ \baar{|l|l|l|l|l|l|l|l|l|} \hline
equation & d x^{1} & d x^{2} & d L_{121} & d L_{122} & d P_{1211} 
& d P_{1212} & d P_{1221} & d P_{1222}\\ \hline
  & & & & & & & & \\
d f^{(R)}_{1212} & \alpha_{12121} & \alpha_{12122} & 2(\G^{1}_{12}
-\G^{2}_{22}) & -2(\G^{1}_{11}+\G^{2}_{12}) & 0 & -2 & 2 &
0\\
{\theta}^{1} & -P_{1211} & -P_{1212} & 1 & 0 & 0 & 0 & 0
& 0\\
{\theta}^{2} & -P_{1221} & -P_{1222} & 0 & 1 & 0 & 0 & 0
& 0\\
2(U\rfloor d {\theta}^{1}) & -U_{1211} & -U_{1212} & 0 & 0 &
U^{1} & U^{2} & 0 & 0\\
2(U\rfloor d {\theta}^{2}) & -U_{1221} & -U_{1222} & 0 & 0 &
0 & 0 & U^{1} & U^{2}\\ 
2(V\rfloor d {\theta}^{1}) & -V_{1211} & -V_{1212} & 0 & 0 &
V^{1} & V^{2} & 0 & 0\\
2(V\rfloor d {\theta}^{2}) & -V_{1221} & -V_{1222} & 0 & 0 &
0 & 0 & V^{1} & V^{2} \\ \hline
\eaar \: . \]
Here, $U$ and $V$ are the first and second Vessiot vector fields 
to be chosen. From the first 5 rows and their reduced counterparts of 
the above Matrix we determine $s_{1}$ and $s_{1}'$. We can easily see that 
the rank of this 5x8-Matrix is 5 and the rank for the reduced matrix, where we
just omit the first 2 columns for $d x^{1}$ and $d x^{2}$, is 5 as well.
Therefore, the Cartan character $s_{1},s_{1}'$ are $s_{1}=s_{1}'=5-3=2$. 
Apart from the same conditions that already hold for $U$, the second Vessiot
vector field $V$ must also satisfy the additional conditions
\be
V^{e}U_{abce}-U^{e}V_{abce} = 0 \: .
\ee
For $V_{1212}$, the same conditions apply as for $U_{1212}$ so that
\be
V_{1212} = V_{1221} + \delta_{1}V^{1} + \delta_{2}V^{2} \label{Ssame}
\ee
with $\delta_{1},\delta_{2}$ as defined above. The only free
components left for $V$ are $V^{1},V^{2},$\\
$V_{1221}$. In order to obtain the rank of the second polar element, we need to
consider all rows of the above matrix. First, we look at the
reduced 6x7-matrix therefore omitting the first two columns under
$d x^{1}$ and $d x^{2}$. Let us denote the $i^{th}$ row by $(i), \: i=1,\cdots
,7$ and check whether the 7 rows are linearly dependent by looking for possible
linear relations among them. We find
\be
(1) = A_{1}\cdot(2) + A_{2}\cdot(3) + B_{1}\cdot(4) + B_{2}\cdot(5) 
+ C_{1}\cdot(6) + C_{2}\cdot(7)\: ,
\ee
where
\[ \baar{ll}
A_{1} = 2(\G^{1}_{12}-\G^{2}_{22})\: , & 
A_{2} =-2(\G^{1}_{11}-\G^{2}_{12})\: ,\\
B_{1} = 2\frac{V^{1}}{\delta}\: , & B_{2} = 2\frac{V^{2}}{\delta}\: ,\\
C_{1} =-2\frac{U^{1}}{\delta}\: , & C_{2} =-2\frac{U^{2}}{\delta}\: ,
\eaar \]
and $\delta=U^{1}V^{2}-U^{2}V^{1}$. This leads to the result 
$s_{2}'$=$6-3-2=1$. Now, we must check whether or not the same linear 
combination occurs amongst the 7 rows for the full polar system:
\be
(1) = A_{1}\cdot(2) + A_{2}\cdot(3) + B_{1}\cdot(4) + B_{2}\cdot(5) 
+ C_{1}\cdot(6) + C_{2}\cdot(7)\: , \label{fliff}
\ee
which must also hold for the {\bf first two columns}. This leads to the
two conditions
\bear
\alpha_{12121} & = & -A_{1}\cdot P_{1211} -A_{2}\cdot P_{1221} 
-B_{1}\cdot U_{1211}-B_{2}\cdot U_{1221} -C_{1}\cdot V_{1211} \nn\\
& & -C_{2}\cdot V_{1221} \: , \nn\\
\alpha_{12122} & = & -A_{1}\cdot P_{1212} -A_{2}\cdot P_{1222} 
-B_{1}\cdot U_{1212}-B_{2}\cdot U_{1222} -C_{1}\cdot V_{1212} \nn\\
& & -C_{2}\cdot V_{1222} \: , \label{fluff}
\eear
where we have to insert the above known expressions for $U_{1212},V_{1212},
V_{1211}$ and $ V_{1222}$ and it is not permitted to restrict 
any of the $U^{e},V^{e}$. By inserting all the above determined values for 
the $A_{i},B_{i},C_{i}$, we see that in the linear relations (\ref{fluff}) the
free components $U_{1211},U_{1221},U_{1222},V_{1221}$ cancel out and both 
relations in (\ref{fluff}) hold identically without imposing any further 
restrictions. Therefore, the same linear combination (\ref{fliff})
holds for the full system. The second character and its reduced 
counterpart therefore coincide so that $s_{2}=s_{2}'=1$ and the involutive 
genus $g$ is $g=\sum_{i=1}^{2}s_{i}=2$.

We can also compute the characters using the tableau and by determining the
torsion we then find out whether the system possesses integrability conditions.
Given a Pfaffian system $\mathcal{P}$ consisting of $s$ 1-forms 
$\theta^{\alpha}$ with independence condition $\Omega = \omega^{1} \wedge
\cdots \wedge \omega^{3}$, we denote by $\pi^{\lambda}$ all the extra forms 
such that $(\theta^{\alpha},\omega^{i},\pi^{\lambda}), \: 1 \leq \alpha 
\leq s$, form a coframe on our formally $N$-dimensional manifold 
$\mathcal{M}$. We can then write
\be
d \theta^{\alpha} = A^{\alpha}_{\lambda i}\wedge \omega^{i} +\frac{1}{2}
B^{\alpha}_{ij}\omega^{i} \wedge \omega^{j} +\frac{1}{2}C^{\alpha}_{\lambda 
\kappa}\pi^{\lambda} \wedge \pi^{\kappa} \: (\mbox{mod} \: \mathcal{I}
(\mathcal{P})) \: . \label{teta}
\ee
In equations (\ref{teta}), the $A^{\alpha}_{\lambda i}$ form the 
{\bf tableau matrix} and the $B^{\alpha}_{ij}$ are called the 
{\bf torsion terms}. Note that if the coefficients $C^{\alpha}_{\lambda 
\kappa}=0$, then the system is said to be {\bf quasi-linear}. In order to 
form a complete coframe for our example (\ref{5REDS2}), we see that we have 
to add three 1-forms $\pi^{\Lambda}$, where $\Lambda$ now is a {\it collective
index} with $\Lambda \in \{ 1211, 1221, 1222 \}$, to the five 1-forms 
$\theta^{1},\theta^{2},\theta^{3},\omega^{1},\omega^{2}$ and so we choose
$$\pi_{1211} := d P_{1211}, \: \: \pi_{1221}:= d P_{1221}, 
\: \: \pi_{1222}:=d P_{1222}\: .$$
The $d P_{1212}$ can be expressed through $\theta^{1}$ as
\bear
d P_{1212} & = & \pi_{1221}
+(\frac{1}{2}\alpha_{12121}+(\G^{1}_{12}-G^{2}_{22})(P_{1211}
-(\G^{1}_{11}+\G^{2}_{12})P_{1221})\omega^{1} +(\frac{1}{2} \nn\\
& & \alpha_{12122}+(\G^{1}_{12}-G^{2}_{22})(P_{1212}
-(\G^{1}_{11}+\G^{2}_{12})P_{1222})\omega^{2}
\: ( \mbox{mod} \: \{\theta^{\alpha} \} ) \: .
\eear
Having accomplished this, we can then write the exterior derivatives of 
the contact conditions as
\bear
d \theta^{1} & \equiv & 0 \nn\\
d \theta^{2} & \equiv & 
- \pi_{1211} \wedge \omega^{1} - \pi_{1221}\wedge \omega^{2} 
+(\frac{1}{2}\alpha_{12121}+(\G^{1}_{12}-\G^{2}_{22})P_{1211} \nn\\
& & -(\G^{1}_{11}+\G^{2}_{12})P_{1221})\omega^{1}\wedge \omega^{2} \nn\\
d \theta^{3} & \equiv & - \pi_{1221}\wedge \omega^{1}-\pi_{1222}\wedge 
\omega^{2} \: .
\eear
Therefore, we obtain for the tableau matrices
\[ \baar{ll} 
A^{\alpha}_{\Lambda 1} = \left( \baar{rrr}
0 & 0 & 0 \nn\\
-1 & 0 & 0 \nn\\
0 & -1 & 0 
\eaar \right) & 
A^{\alpha}_{\Lambda 2} = \left( \baar{rrr}
0 & 0 & 0 \nn\\
0 & -1 & 0 \nn\\
0 & 0 & -1 
\eaar \right) \: , \eaar \]
where $\alpha =1,2,3$ and $\Lambda$ is one of the collective indices 
$\Lambda \in \{ 1211, 1221, 1222 \}$. This leads to $s_{1}'=2, \: 
s_{2}'=1$, and, the only non-vanishing torsion term is given by 
\be
B^{2}_{12} = -[\frac{1}{2}\alpha_{12121}-(\G^{1}_{12}-\G^{2}_{22})P_{1211}
+(\G^{1}_{11}+\G^{2}_{12})P_{1221}] \: . 
\ee
If we wish to absorb the torsion coefficients, we must find a transformation
$\Phi$ with $\pi^{\lambda} \: \rightarrow \: \pi^{\lambda}
+p^{\lambda}_{\: i}\omega^{i}$ and quantities $p^{\lambda}_{i}$ such that
$$0={\tilde{B}}^{\alpha}_{ij}=B^{\alpha}_{ij}+A^{\alpha}_{ \lambda j}
p^{\lambda}_{\: i}-A^{\alpha}_{\lambda i}p^{\lambda}_{\: j} \: .$$
In our case this leads to the system
\bear
0 & = & A^{1}_{\Lambda 2}p^{\Lambda}_{\: 1}-A^{1}_{\Lambda 1}
p^{\Lambda}_{\: 2} \nn\\
0 & = & B^{2}_{12}+A^{2}_{\Lambda 2}p^{\Lambda}_{\: 1}
-A^{2}_{\Lambda 1}p^{\Lambda}_{\: 2} \nn\\
0 & = & A^{3}_{\Lambda 2}p^{\Lambda}_{\: 1}-A^{3}_{\Lambda 1}
p^{\Lambda}_{\: 2} \: . \label{2torsi}
\eear
One solution {\it Ansatz} to fulfil (\ref{2torsi}) is to choose
$ p^{2}_{\: 1}:=B^{2}_{12}$ while all other $p^{\Lambda}_{\: i}$ vanish: thus
the torsion is absorbed and the system is therefore in involution
\cite{Cart,Yang}.

Adding the exterior derivative of the differential gauge condition\\
${L_{12}}^{s}_{\: ;s}=0$, which is locally given by 
\bear
d {L_{12}}^{s}_{\: ;s} & = & g^{11}d P_{1211}+g^{12}(d P_{1221}+d
P_{1212}) +g^{22}d P_{1222}- \delta_{3}d L_{121}- \delta_{4}d L_{122} \nn\\ 
& & + \delta_{5}d x^{1} + \delta_{6}d x^{2} \: ,
\eear
where $\delta_{3}, \cdots , \delta_{6}$ are defined in Appendix A, does
not change the results qualitatively. We can write down the polar systems
again and from this obtain the polar matrices of which the matrix of 
$H(E^{2})_{x}$ is given in Appendix A.

Obviously $s_{0}=s_{0}'=4$ and we find that no linear combinations are 
possible amongst the first 6 rows or their reduced counterparts so that\\
$s_{1}=s_{1}'=2$. But for $s_{2}'$ there exist multipliers $A_{i},B_{j}$ 
such that
\bear
(7) & = &
A_{1}\cdot (1)+A_{2}\cdot (2)+A_{3}\cdot (3)+A_{4}\cdot (4)+A_{5}\cdot (5)
+A_{6}\cdot (6)\: ,\nn\\\
(8) & = &
B_{1}\cdot (1)+B_{2}\cdot (2)+B_{3}\cdot (3)+B_{4}\cdot (4)+B_{5}\cdot (5)
+B_{6}\cdot (6)
\eear
now with multipliers $A_{1}, \cdots, A_{6}$ and $B_{1}, \cdots , B_{6}$ as 
given in Appendix A. In order for these linear combinations to hold for the 
full rows (7) and (8) as well, the following 4 equations have to hold:
\bear
V_{1211} & = & -A_{1}\alpha_{12121}-A_{2}\delta_{5}+A_{3}P_{1211}
               +A_{4}P_{1221}+A_{5}U_{1211}+A_{6} \nn\\
& & U_{1221}, \label{5Inv1}\\
V_{1221} & = & -B_{1}\alpha_{12121}-B_{2}\delta_{5}+B_{3}P_{1211}
               +B_{4}P_{1221}+B_{5}U_{1211}+B_{6} \nn\\
& & U_{1221}, \label{5Inv2}\\
V_{1212} & = & -A_{1}\alpha_{12122}-A_{2}\delta_{6}+A_{3}P_{1212}
               +A_{4}P_{1222}+A_{5}U_{1212}+A_{6} \nn\\
& & U_{1222}, \label{5Inv3}\\
V_{1222} & = & -B_{1}\alpha_{12122}-B_{2}\delta_{6}+B_{3}P_{1212}
               +B_{4}P_{1222}+B_{5}U_{1212}+B_{6} \nn\\
& & U_{1222}, \label{5Inv4}
\eear
where $U$ is the first Vessiot vector field and $V$ the second 
one. After a lengthy calculation of which details are given in \cite{Ger1} 
we find that with (\ref{Ssame}) the above expressions (\ref{5Inv1}) 
to (\ref{5Inv4}) hold. Therefore, the full rows (7) and (8) are again
linearly dependent which leads to $s_{2}=s_{2}'=0$ and $g=2$. Note that all 
the above results can be verified using a REDUCE code given in \cite{Ger1}.
\subsection{The 2-dimensional Riemann-Lanczos Problem as a System of PDEs}
The Riemann-Lanczos problem in 2 dimensions is an example of a system of 
partial differential equations for which applying Janet-Riquier theory is 
much more economical. We shall also see in Section C below that it is of great 
importance to compute the characters using $\delta$-regular coordinates. 
There, if characteristic coordinates are used, they will lead to non-intrinsic
results so that a coordinate transformation will then have to be 
carried out. We have a system of equations $\mathcal{R}_{1}$ with $n=m=2,q=1$ 
and the symbol $\mathcal{M}_{1}$ consists only of a single 
equation
\be
-2V_{1212} + 2V_{1221} = 0 \: ,
\ee
which is of class 2 whatever the choice of the orderly ranking.
Note that here and in subsequent sections, where Janet-Riquier theory is 
applied, the {\it symbol variables} $V_{abcd},V_{abcde}$ have no connection to
Vessiot vector fields $V$. This yields
$$ \alpha^{(1)}_{1}=2, \quad \alpha^{(2)}_{1} =1 \: .$$
Trivially, all variables are multiplicative variables and the symbol of
one equation is always involutive. But, we must examine whether this 
system is formally integrable. We only need to compute $\mathcal{R}_{2}$ 
and its projection $\mathcal{R}^{(1)}_{1}$ and check whether 
$\mathcal{R}_{2}=\mathcal{R}^{(1)}_{1}$, where $\mbox{dim}(\mathcal{R}_{1})
=5$ and $\mbox{dim}(\mathcal{M}_{1})=3$. Our prolonged system of equations 
$\mathcal{R}_{2}$ consists of 
$f^{(R)}_{1212},f^{(R)}_{1212,1},f^{(R)}_{1212,2},$ namely,
\bear
0 & = & R_{1212}-2P_{1212}+2P_{1221}+2L_{121}(\G^{1}_{12}-\G^{2}_{22})
-2L_{122}(\G^{1}_{11}+\G^{2}_{12}) \nn\\
0 & = & R_{1212,1}-2S_{12121}+2S_{12211}+2P_{1211}
(\G^{1}_{12}-\G^{2}_{22}) \nn\\
& & -2P_{1221}(\G^{1}_{11}+\G^{2}_{12})+2L_{121}(\G^{1}_{12}
-\G^{2}_{22})_{,1}-2L_{122}(\G^{1}_{11}+\G^{2}_{12})_{,1} \nn\\
0 & = & R_{1212,2}-2S_{12122}+2S_{12212}
+2P_{1212}(\G^{1}_{12}-\G^{2}_{22})\nn\\
& & -2P_{1222}(\G^{1}_{11}+\G^{2}_{12})+2L_{121}(\G^{1}_{12}-\G^{2}_{22})_{,2}
-2L_{122}(\G^{1}_{11}+\G^{2}_{12})_{,2} \label{5Symp2} \: .
\eear
No integrability conditions can be created because
$$\mbox{dim}(\mathcal{R}^{(1)}_{1})=\mbox{dim}(\mathcal{R}_{2})-\mbox{dim}
(\mathcal{M}_{2})=9-4=5=\mbox{dim}(\mathcal{R}_{1})$$
and the system is linear so that formal integrability has been shown and 
the system $\mathcal{R}_{1}$ is involutive. 

If we incorporate the differential gauge condition ${L_{ab}}^{s}_{\: ;s}=0$,
then $\mathcal{M}_{1}$ is formed by 
\bear
-2V_{1212} + 2V_{1221} & = & 0\nn\\
g^{11}V_{1211} +g^{12}V_{1221}+ g^{12}V_{1212}+g^{22}V_{1222} & = & 0
\: ,
\eear
which has two equations of class 2 again whatever the choice of the orderly 
ranking be and we obtain $\alpha^{(1)}_{1}=2, \: \alpha^{(2)}_{1} =0\: .$
In order to check whether $\mathcal{M}_{1}$ is involutive, we prolong 
to the corresponding $\mathcal{M}_{2}$ which is formed by 
\bear
-2 V_{12112}+ 2 V_{12211} & = & 0\nn\\
-2 V_{12122}+ 2 V_{12212} & = & 0\nn\\
g^{11}V_{12111} +g^{12}V_{12211}+ g^{12}V_{12112}
+g^{22}V_{12212} & = & 0\nn\\
g^{11}V_{12112} +g^{12}V_{12212}+ g^{12}V_{12122}
+g^{22}V_{12222} & = & 0\: ,
\eear
We get $r(\mathcal{M}_{2})=4= k \cdot \beta^{(k)}_{1}$ and therefore 
$\mathcal{M}_{1}$ is involutive. 
We must check that the new system $\mathcal{R}_{1}$ is formally 
integrable again and so add the following 3 equations 
${L_{12}}^{s}_{\: ;s},\: \pr_{1}{L_{12}}^{s}_{\: ;s},
\: \pr_{2}{L_{12}}^{s}_{\: ;s}$ to (\ref{5Symp2}):
\bear
0 & = & g^{11}P_{1211}+g^{12}(P_{1221}+P_{1212})+g^{22}P_{1222}
-L_{121}\delta_{3}-L_{122}\delta_{4} \nn\\
0 & = & g^{11}S_{12111}+g^{12}(S_{12211}+S_{12112})+g^{22}S_{12212}
+g^{11}_{\: ,1}P_{1211}+g^{12}_{\: ,1}(P_{1221}+P_{1212}) \nn\\
& & +g^{22}_{\: ,1}P_{1222}
-P_{1211}\delta_{3}-P_{1221}\delta_{4}-L_{121}\delta_{3,1}
-L_{122}\delta_{4,1} \nn\\
0 & = & g^{11}S_{12112}+g^{12}(S_{12212}+S_{12122})+g^{22}S_{12222}
+g^{11}_{\: ,2}P_{1211}+g^{12}_{\: ,2}(P_{1221}+P_{1212}) \nn\\
& & +g^{22}_{\: ,2}P_{1222}
-P_{1212}\delta_{3}-P_{1222}\delta_{4}-L_{121}\delta_{3,2}
-L_{122}\delta_{4,2} \: , \label{5Sympdg2}
\eear
where $\delta_{3},\delta_{4}$ are as defined in Appendix A. The system 
(\ref{5Symp2}) together with (\ref{5Sympdg2}) now forms the prolonged
system of PDEs $\mathcal{R}_{2}$ including the differential gauge 
condition and its partial derivatives. Again, we cannot solve any of the
4 second-order PDEs in such a way that no second-order derivatives
remain. In order to show this, we can use the symbol matrix of 
$\mathcal{M}_{2}$ as well:
\[ \baar{|c|c|c|c|c|c|} \hline
V_{12111} & V_{12112} & V_{12122} & V_{12211} & V_{12212} &
V_{12222} \nn\\ \hline
0 & -2 & 0 & 2 & 0 & 0 \nn\\
0 & 0 & -2 & 0 & 2 & 0 \nn\\
g^{11} & g^{12} & 0 & g^{12} & g^{22} & 0 \nn\\
0 & g^{11} & g^{12} & 0 & g^{12} & g^{22} \\ \hline
\eaar \: , \]
which has maximal rank $r(\mathcal{M}_{2})=4$ and no integrability 
conditions can arise because of the linearity of the system. This means that
$\mathcal{R}^{(1)}_{1}=\mathcal{R}_{1}$, thus making $\mathcal{R}_{1}$ 
formally integrable and involutive. We conclude:
\bt{Proposition}\\
In any two-dimensional spacetime, the Riemann-Lanczos equations are
in involution with respect to the spacetime coordinates and the 
differential gauge condition does not affect the result. The corresponding 
Cartan characters $(s_{0},s_{1},s_{2})$ for the involutory systems are (3,2,1)
when no differential gauge condition is imposed but (4,2,0) when the 
differential gauge condition is imposed.
\et
\subsection{General Solution to the Riemann-Lanczos Problem in 2 Dimensions
and Characteristic Coordinates}
It is instructive to repeat some of the above calculations using different
local coordinates. First, we look at the general 2-dimensional spacetime with 
Lorentzian signature and line element in characteristic coordinates 
$x^{1},x^{2}$ written as
$$d s^{2}= -e^{2\rho}d x^{1}d x^{2} \: ,$$ 
where $\rho$ is an arbitrary function of $x^{1},x^{2}$. The only two 
equations we have to solve are $f^{(R)}_{1212}=0,{L_{12}}^{s}_{;s}=0:$
\bear
0 & = &
-e^{2\rho}{\rho}_{,x^{1}x^{2}}+4L_{121}{\rho}_{,x^{2}}-4L_{122}
{\rho}_{,x^{1}}-2P_{1212}+2P_{1221} \: ,\nn\\
0 & = & 2e^{-2\rho}(2L_{121}{\rho}_{,x^{2}}+2L_{122}{\rho}_{,x^{1}}
        -P_{1212}-P_{1221}) \label{5dims2} \: .
\eear
We find that the general solution is of the form
\bear 
L_{121} & = & e^{2\rho}f_{1}(x^{1})-\frac{1}{4}e^{2\rho}\rho_{,x^{1}}\: ,\nn\\
L_{122} & = & e^{2\rho}f_{2}(x^{2})+\frac{1}{4}e^{2\rho}\rho_{,x^{2}}\: ,
\eear
where $f_{1}(x^{1}), f_{2}(x^{2})$ are 2 arbitrary functions depending
on one local coordinate each. This is in agreement with the
result for the Cartan characters claiming that the general solution 
depends on two arbitrary functions of one variable each. But can 
we obtain the intrinsic values for the characters using the 
above coordinate frame? The symbol derived from (\ref{5dims2}) is 
\bear
0 & = & -2 V_{1212}+2V_{1221} \nn\\
0 & = & -V_{1212}-V_{1221} \: , \label{2unno}
\eear
from which we obtain $\beta^{(1)}_{1}=\beta^{(2)}_{1}=1$ based on the 
ranking stemming from $x^{2} \succ x^{1}$. This leads to $\alpha^{(1)}_{1}=
\alpha^{(2)}_{1}=1$, which {\bf cannot be} the intrinsic result as we already
solved the 2-dimensional problem in the previous section. We see that in 
(\ref{2unno}) only one variable of class $2$, namely $V_{1212}$, and only 
one variable of class $1$, namely $V_{1221}$, occurs whereas neither $V_{1211}$
nor $V_{1222}$ occur at all to make both equations be of class 
$2$. As this can happen in another coordinate system as we shall see for the 
3-dimensional case below as well, the above coordinate frame is not 
$\delta$-regular and we need to perform a coordinate transformation of the form
\bear
d \tilde{x}^{1} & = & a_{11}d x^{1} + a_{12} d x^{2} \nn\\
d \tilde{x}^{2} & = & a_{21}d x^{1} + a_{22} d x^{2} \label{5cord2}
\eear
in order to obtain the correct values for $\alpha^{(1)}_{1},\alpha^{(2)}_{1}$. 
After such a transformation, the new symbol in orthonomic form is
\bear
V_{1222} & = & -\frac{a_{21}}{a_{22}}V_{1221} \nn\\
V_{1212} & = & -\frac{a_{11}}{a_{12}}V_{1211} \: .
\eear
Now, both equations are of class 2 so that $\beta^{(1)}_{1}=0, 
\beta^{(2)}_{1}=2$ which produces $\alpha^{(1)}_{1}=2, \alpha^{(2)}_{1}=0$ 
which is the known intrinsic result for the Cartan characters. 

For spaces with Euclidean signature we can write their line element as
$$d s^{2}= e^{2\rho}({(d x^{1})}^{2}+{(d x^{2})}^{2}) \: ,$$ 
where $\rho$ is again an arbitrary function of $x^{1},x^{2}$ leading 
to the system of equations
\bear
0 & = &
-e^{2\rho}({\rho}_{,x^{1}x^{1}}+{\rho}_{,x^{2}x^{2}})+4L_{121}{\rho}_{,x^{2}}
-4L_{122}{\rho}_{,x^{1}}-2P_{1212}+2P_{1221} \: ,\nn\\
0 & = & e^{-2\rho}(-2L_{121}{\rho}_{,x^{1}}-2L_{122}{\rho}_{,x^{2}}
        +P_{1211}+P_{1222}) \label{5dims22} \: .
\eear
The solution to (\ref{5dims22}) resembles the one for (\ref{5dims2}) and is
\bear
L_{121} & = & e^{2\rho}(f_{1}(x^{1}+x^{2})-\frac{1}{2}\rho_{,x^{2}}) \nn\\
L_{122} & = & e^{2\rho}(f_{2}(x^{1}-x^{2})+\frac{1}{2}\rho_{,x^{1}}) \: .
\eear
In this case we find the intrinsic values for the characters to be 
$\alpha^{(1)}_{1}=2,$\\
$\alpha^{(2)}_{1}=0$ using (\ref{5dims22}) directly
due to the fact that here we did not use characteristic local coordinates.
\section{The Riemann-Lanczos Problem in 3 Dimensions: Involution or 
Prolongation?}
In this section we shall discuss the Riemann-Lanczos problem in 3 dimensions
as an EDS and as a Janet-Riquier system of partial differential equations 
(PDEs). First, we shall use EDS theory to find out more about the problem.
\subsection{The 3-dimensional Riemann-Lanczos Problem as an EDS}
In 3 spacetime dimensions with local coordinates $x^{1},x^{2},x^{3}$, 
we obtain 8 independent components of the Lanczos tensor, namely, 
$L_{121},L_{122},L_{131},L_{133},$\\
$L_{232},L_{233},L_{123},L_{132}$ when imposing the cyclic conditions
(\ref{1cyclic}). Each of them has three first-order partial
derivatives so that we use the jet bundle $\mathcal{J}^{1}(\mathbb{R}^{3},
\mathbb{R}^{8})$ with formal dimension $N=35$ to express the EDS.
We obtain the 6 independent Riemann-Lanczos equations whose 
exterior derivatives are $d f^{(R)}_{1212}, d f^{(R)}_{1313},$\\ 
$d f^{(R)}_{2323}, d f^{(R)}_{1213}, d f^{(R)}_{1223}, d f^{(R)}_{1323}$.
These latter equations can locally be written as 
\bear
0 & = & -2d P_{1212}+2d P_{1221}+d L_{121}(2\G^{1}_{12}+2\G^{2}_{22})
-d L_{122}(2\G^{1}_{11}+2\G^{2}_{12}) +2\G^{3}_{22}\nn\\
& & \cdot d L_{131}+ 2\G^{3}_{11}d L_{232}+2\G^{3}_{12}d L_{123}
-4\G^{3}_{12}d L_{132}+(R_{1212,e}+\alpha_{1212e})d x^{e}\nn\\
0 & = & -2d P_{1313}+2d P_{1331}+d L_{131}(2\G^{1}_{13}+2\G^{3}_{33})
-d L_{133}(2\G^{1}_{11}+\G^{3}_{13}) +2\G^{2}_{33}\nn\\
& & \cdot d L_{121}-2\G^{2}_{11}d L_{233}-4\G^{2}_{13}d L_{123}
+2\G^{2}_{13}d L_{132}+(R_{1313,e}+\alpha_{1313e})d x^{e}\nn\\
0  & = & -2d P_{2323}+2d P_{2332}+2d L_{232}(\G^{2}_{23}+\G^{3}_{33})
-2 d L_{233}(\G^{2}_{22}+\G^{3}_{23}) -2\G^{1}_{33}\nn\\
& & \cdot d L_{122}-2\G^{1}_{22}d L_{133}+2\G^{1}_{23}d L_{123}
+2\G^{1}_{23}d L_{132} +(R_{2323,e}+\alpha_{2323e})d x^{e}\nn\\
0 & = & -d P_{1213}+d P_{1231}-d P_{1312}+d P_{1321}+d L_{121}
(\G^{1}_{13}+2\G^{2}_{23})+d L_{131}\nn\\
& & \cdot (\G^{1}_{12}+2\G^{3}_{23})-\G^{3}_{12}d L_{133}-\G^{2}_{13}d L_{122}
-\G^{2}_{11}d L_{232}+\G^{3}_{11}d L_{233}-d L_{123}\nn\\
& & \cdot (\G^{1}_{11}+2\G^{2}_{12}-\G^{3}_{13})-d L_{132}(2\G^{3}_{13}
+\G^{1}_{11}-\G^{2}_{12})+(R_{1213,e}+\alpha_{1213e})d x^{e}\nn\\
0  & = & -d P_{1223}+2d P_{1232}-d P_{1322}+d P_{2321}+d L_{122}
(2\G^{1}_{13}+\G^{2}_{23})-d L_{232}\nn\\
& & \cdot (\G^{2}_{12}+2\G^{3}_{13})-\G^{1}_{23}d L_{121}+\G^{3}_{12}d L_{233}
+\G^{1}_{22}L_{131}-\G^{3}_{22}d L_{133}-d L_{123}\nn\\
& & \cdot (\G^{1}_{12}+2\G^{2}_{22}+\G^{3}_{23})+d L_{132}(2\G^{3}_{23}
+\G^{2}_{22}-\G^{1}_{12})+(R_{1223,e}+\alpha_{1223e})d x^{e}\nn\\
0 & = & -2d P_{1323}+d P_{1332}+d P_{1233}+d P_{2331}-d L_{133}
(2\G^{1}_{12}+\G^{3}_{23})-d L_{233}\nn\\
& & \cdot (2\G^{2}_{12}+\G^{3}_{13})+\G^{2}_{13}d L_{232}+\G^{2}_{33}d L_{122}
+\G^{1}_{23}L_{131}-\G^{1}_{33}d L_{121}-d L_{123}(2\G^{2}_{23} \nn\\
& & -\G^{3}_{33}-\G^{1}_{13})+d L_{132}(\G^{1}_{13}
+2\G^{3}_{33}+\G^{2}_{23})+(R_{1323,e}+\alpha_{1323e})d x^{e} \: , 
\label{5Pudel}
\eear
where $\alpha_{abcde}$ is defined in Appendix B. The polar matrix
consists of 35 columns for the $(d x^{e}, d L_{abc}, d P_{abcd})$ and 
we split the first 6 rows stemming from (\ref{5Pudel}) into 3 blocks 
which we call $( M_{01} \quad M_{02} \quad M_{03} )$ so that the full
polar matrix of $H(E^{3})_{x}$ is given by
\[ \left( \baar{lll}
M_{01} & M_{02} & M_{03} \nn\\
M(P) & 1 & 0 \nn\\
M(U) & 0 & N(U) \nn\\
M(V) & 0 & N(V) \nn\\
M(Z) & 0 & N(Z) 
\eaar \right) \: , \] 
where we denote the truncated matrix obtained by leaving out the rows 
$(M_{01} \:\: M_{02} \:\: M_{03})$ by $M_{1}$. All the matrix expressions can 
be found in Appendix B and a more detailed explanation is given in 
\cite{Ger1}. The 
first 8 rows from $M_{1}$ together with $(M_{01} \: M_{02} \: M_{03})$ 
lead to $s_{0}=s_{0}'=14$. Then, for $s_{1}$ and $s_{1}'$ there is 
neither a linear combination of the full polar system nor one of the 
reduced polar system possible. We have $s_{1}=s_{1}'=8$ because the 
rank of the relevant 22x35-matrix consisting of 
$(M_{01} \quad M_{02} \quad M_{03})$ and the first 16 rows of $M_{1}$
is $22=s_{0}+s_{1}=s_{0}'+s_{1}'$ and so $s_{0}=s_{0}'=8$. But now, for the 
second reduced polar system, we find that there is exactly one linear 
combination possible which we can write down as 
\bear
0 & = & A_{1}\cdot (row_{1})+A_{2}\cdot (row_{2}) + \cdots + A_{6}\cdot
(row_{6}) \nn\\
& & B_{1}\cdot (row_{7})+B_{2}\cdot (row_{8}) + \cdots + B_{8}\cdot
(row_{14}) \nn\\
& & +C_{1}\cdot (row_{15})+C_{2}\cdot (row_{16}) + \cdots + C_{8}\cdot
(row_{22}) \: , \label{5LinC}
\eear
where the linear multipliers are given in Appendix B.
If we wish that this linear combination for the full polar system
be valid, we have to claim that the 3 equations (for $e=1,2,3$) hold
\bear
0 & = & {\tilde{\alpha}}_{1212e}A_{1} + {\tilde{\alpha}}_{1313e}A_{2}
+{\tilde{\alpha}}_{2323e}A_{3} + {\tilde{\alpha}}_{1213e}A_{4}
+{\tilde{\alpha}}_{1223e}A_{5} + {\tilde{\alpha}}_{1323e}A_{6}\\
 & & -U_{121e}B_{1}-U_{122e}B_{2}-U_{131e}B_{3}-U_{133e}B_{4}
-U_{232e}B_{5}-U_{233e}B_{6} \nn\\
& & -U_{123e}B_{7}-U_{132e}B_{8}-V_{121e}C_{1}-V_{122e}C_{2}-V_{131e}C_{3}
-V_{133e}C_{4} \nn\\
& & -V_{232e}C_{5}-V_{233e}C_{6}-V_{123e}C_{7}-V_{132e}C_{8} \: , \nn 
\eear
where ${\tilde{\alpha}}_{abcde}$ is as defined in Appendix B,
without imposing any further restrictions on the first two Vessiot vector 
fields $U,V$. This can easily be checked by looking at how many of the
$U_{abcd}$ and how many of the $V_{abcd}$ can be chosen arbitrarily.
For the 24 $U_{abcd}$ there are 6 restrictions, so 18 can be chosen 
arbitrarily. For the $V_{abcd}$ we have 14 restrictions, so 10 remain
arbitrary. Based on this argument, we find that some of the $U_{abcd},
V_{abcd}$ never cancel out and the linear combination 
(\ref{5LinC}) does not hold for the full system so that $s_{2}
=s_{2}'+1$ such that $s_{2}'=7$ and $s_{2}=8$ \cite{Ger1}.
This system is clearly not in involution and its reduced characters 
$(s_{0}',s_{1}',s_{2}',s_{3}')$ are $(14,8,7,3)$.
We can test this result with the REDUCE code given in 
\cite{Ger1} and immediately obtain (8,7,3) for the reduced 
characters $(s_{1}',s_{2}',s_{3}')$ and that the system is not in involution.
In order to carry out this computation it is sufficient to use an arbitrary 
diagonalized line element which describes 3 -dimensional spacetimes 
completely as shown in \cite{Turk}.

We can also use the tableau matrix to calculate this result as shown next.
First, we must complete our set of 14 one-forms together with the 3 
$\omega^{e}$ to a complete coframe consisting of $N=35$ elements by 
introducing the 18 $\pi^{\Lambda}$ such that $(d f^{(R)}_{abcd},K_{abc},
\omega^{e},\pi^{\Lambda})$ forms a cobasis on $\mathcal{M}$. We introduce the 
18 $\pi^{\Lambda}$ using collective indices $\Lambda$ with $\Lambda =1, \cdots
,18$ with the following correspondence between $\pi^{\Lambda} \leftrightarrow 
d P_{abcd}$:
\[ \baar{lll}
\pi^{1} \leftrightarrow d P_{1211} \: , \quad & 
\pi^{2} \leftrightarrow d P_{1221} \: , \quad & 
\pi^{3} \leftrightarrow d P_{1311} \: , \nn\\
\pi^{4} \leftrightarrow d P_{1331} \: , \quad & 
\pi^{5} \leftrightarrow d P_{2321} \: , \quad & 
\pi^{6} \leftrightarrow d P_{2331} \: , \nn\\
\pi^{7} \leftrightarrow d P_{1231} \: , \quad & 
\pi^{8} \leftrightarrow d P_{1321} \: , \quad & 
\pi^{9} \leftrightarrow d P_{1222} \: , \nn\\
\pi^{10} \leftrightarrow d P_{1312} \: , \quad & 
\pi^{11} \leftrightarrow d P_{1332} \: , \quad & 
\pi^{12} \leftrightarrow d P_{2322} \: , \nn\\
\pi^{13} \leftrightarrow d P_{2332} \: , \quad & 
\pi^{14} \leftrightarrow d P_{1232} \: , \quad & 
\pi^{15} \leftrightarrow d P_{1322} \: , \nn\\
\pi^{16} \leftrightarrow d P_{1333} \: , \quad & 
\pi^{17} \leftrightarrow d P_{2333} \: , \quad & 
\pi^{18} \leftrightarrow d P_{1233} \: . 
\eaar \]
When using the correspondence $\theta^{\alpha} \leftrightarrow K_{abc}$ based 
on equations (18) in \cite{DoGe1} the 8 $d K_{abc}$ can be recast as
\bear
d \theta^{1} & \equiv & -\pi^{1}\wedge \omega^{1}-\pi^{2}\wedge \omega^{2}
-\pi^{7}\wedge\omega^{3}+\pi^{10}\wedge\omega^{3}-\pi^{8}\wedge\omega^{3} \nn\\
& & +\frac{1}{2}B^{1}_{i2}\omega^{i}\wedge\omega^{2}+\frac{1}{2}B^{1}_{i3}
\omega^{i}\wedge\omega^{3} \nn\\
d \theta^{2} & \equiv & -\pi^{2}\wedge \omega^{1}-\pi^{9}\wedge \omega^{2}
-2\pi^{14}\wedge\omega^{3}-\pi^{5}\wedge\omega^{3}+\frac{1}{2}B^{2}_{i3}
\omega^{i}\wedge\omega^{3} \nn\\
d \theta^{3} & \equiv & -\pi^{3}\wedge \omega^{1}-\pi^{10}\wedge\omega^{2}
-\pi^{4}\wedge\omega^{3}+\frac{1}{2}B^{3}_{i3}\omega^{i}\wedge\omega^{3} \nn\\
d \theta^{4} & \equiv & -\pi^{4}\wedge \omega^{1}-\pi^{11}\wedge \omega^{2}
-\pi^{16}\wedge\omega^{3} \nn\\
d \theta^{5} & \equiv & -\pi^{5}\wedge \omega^{1}-\pi^{12}\wedge \omega^{2}
-\pi^{13}\wedge\omega^{3}+\frac{1}{2}B^{5}_{i3}\omega^{i}\wedge\omega^{3} \nn\\
d \theta^{6} & \equiv & -\pi^{6}\wedge \omega^{1}-\pi^{13}\wedge\omega^{2}
-\pi^{17}\wedge\omega^{3} \nn\\
d \theta^{7} & \equiv & -\pi^{7}\wedge\omega^{1}-\pi^{14}\wedge\omega^{2}
-\pi^{18}\wedge\omega^{3} \nn\\
d \theta^{8} & \equiv & -\pi^{8}\wedge\omega^{1}-\pi^{15}\wedge\omega^{2}
-\frac{1}{2}\pi^{11}\wedge\omega^{3}-\frac{1}{2}\pi^{8}\wedge\omega^{3}
-\frac{1}{2}\pi^{6}\wedge\omega^{3} \nn\\
& & +\frac{1}{2}B^{8}_{i3}\omega^{i}\wedge\omega^{3} \label{8KDES} \: .
\eear
We can easily calculate the reduced characters when using the tableau matrices
derived from (\ref{8KDES}) and obtain $s_{1}'=8$, $s_{2}'=7$ and $s_{3}'=3$.
We also find that all remaining torsion coefficients $B^{\alpha}_{ij}$ in 
(\ref{8KDES}) can be absorbed so that no integrability conditions can occur 
for the system (\ref{5Pudel}) itself. However, because we have the rank 
deficiency in the reduced polar matrix leading to $s_{2}=s_{2}'+1$, we must 
prolongate the system to second order. We introduce new jet variables 
$S_{abcde}$, of which the projection back onto the spacetime manifold 
corresponds to $S_{abcde}:=P_{abcd,e}=L_{abc,de}$, and denote the 24 contact 
conditions arising from them by $K_{abcd}$. The prolonged Pfaffian system 
then is given by
\bear
0 & = & f^{(R)}_{abcd} \nn\\
0 & = & f^{(R)}_{abcd,e} \nn\\
0 & = & d f^{(R)}_{abcd} \nn\\
0 & = & d f^{(R)}_{abcd,e} \nn\\
K_{abc} & = & d L_{abc}-P_{abcd}d x^{e} \nn\\
K_{abcd} & = & d P_{abcd}-S_{abcde}d x^{e} \label{Edscheck}
\eear
together with the exterior derivatives $d K_{abc}$ and $d K_{abcd}$. 
Calculations using an adapted REDUCE computer code suggest that the system
(\ref{Edscheck}) is not in involution and integrability conditions occur.

Note that instead of the prolongation to (\ref{Edscheck}) we can carry out the
same prolongation which was carried out for the Riemann-Lanczos problem in 4 
dimensions in \cite{Bam2}. This may lead an involutory system eventually but we
leave this as future work. However, the process of prolongation often becomes
simpler when Janet-Riquier theory is used what we are going to next.
\subsection{The 3-dimensional Riemann-Lanczos Problem as a System of PDEs}
The symbol $\mathcal{M}_{1}$ for the unprolonged Riemann-Lanczos problem in 3
dimensions consists of the 6 linear equations derived from $f^{(R)}_{1212}, 
f^{(R)}_{1313}, f^{(R)}_{2323}, f^{(R)}_{1213},$\\
$f^{(R)}_{1223}, f^{(R)}_{1323}$ and is given in orthonomic form as
\[ \baar{lr}
V_{1212} = V_{1221}  \qquad & x^{1} \: x^{2} \: \bullet \nn\\
 & \nn\\
V_{1313} = V_{1331} \qquad & x^{1} \: x^{2} \: x^{3} \nn\\
 & \nn\\
V_{2323} = V_{2332} \qquad & x^{1} \: x^{2} \: x^{3} \nn\\
 & \nn\\
V_{1213} = V_{1231}-V_{1312}+V_{1321} \qquad & 
x^{1} \: x^{2} \: x^{3} \nn\\
 & \nn\\
V_{1223} = 2V_{1232}-V_{1322}+V_{2321} \qquad &
x^{1} \: x^{2} \: x^{3} \nn\\
 & \nn\\
V_{1323}= \frac{1}{2}(V_{1332}+V_{1233}+V_{2331}) \qquad &
x^{1} \: x^{2} \: x^{3} \label{5Rie3Sym} .
\eaar \]
We imposed the ranking $x^{3} \succ x^{2} \succ x^{1}$ which then induced the
ranking $P_{abc3} \succ P_{abc2} \succ P_{abc1}$ and amongst each set 
$P_{abce}$ the ranking $P_{233e} \succ P_{232e} \succ P_{132e} \succ P_{123e}
\succ P_{133e} \succ P_{131e} \succ P_{122e} \succ P_{121e}$. From this we 
now obtain  $\beta^{(1)}_{1}=0,\beta^{(2)}_{1}=1,\beta^{(3)}_{1}=5$ and 
$\alpha^{(1)}_{1}=8,\alpha^{(2)}_{1}=7,\alpha^{(3)}_{1}=3$. But, we must
check whether we actually used $\delta$-regular coordinates and
we perform the following coordinate transformation
\bear
d \tilde{x}^{1} & = & 
a_{11} d x^{1}+a_{12} d x^{2}+a_{13} d x^{3}\nn\\
d \tilde{x}^{2} & = & 
a_{21} d x^{1}+a_{22}d x^{2}+a_{23} d x^{3}\nn\\
d \tilde{x}^{3} & = & a_{31}d x^{1}+a_{32} d x^{2}+a_{33}d x^{3} \: , 
\label{5delta3}
\eear
which leaves us with the symbol in orthonomic form in the new coordinates
\bear
V_{1223} & = & \frac{1}{a_{13}}(B_{1}+a_{23}V_{1213}) \nn\\
V_{1333} & = & \frac{1}{a_{13}}(B_{2}+a_{33}V_{1313}) \nn\\
V_{2333} & = & \frac{1}{a_{23}}(B_{3}+\frac{a_{33}}{a_{13}}(B_{5}
+\frac{a_{33}}{a_{13}}B_{1}+\frac{a_{23}}{a_{13}}B_{4} \nn\\
& & +2V_{1213}\frac{a_{33}a_{23}}{a_{13}}-3a_{23}V_{1233}
+\frac{a^{2}_{23}}{a_{13}}V_{1313}) \nn\\
V_{1323} & = & \frac{1}{a_{13}}(B_{4}+a_{33}V_{1213}-a_{13}V_{1233}
+a_{23}V_{1313}) \nn\\
V_{2332} & = & f(V_{abc1}, V_{abc2}) 
\: . \label{5Rie3del}
\eear
In (\ref{5Rie3del}) we dropped the ${~}$ and the $B_{i}$ and $f(V_{abc1},
V_{abc2})$ are given in Appendix C. We see from (\ref{5Rie3del}) and Appendix
C that no variable of the form $V_{abc3}$ remains in the last equation for 
any choice of the $a_{ij}$ in (\ref{5delta3}) so that we obtain 5 equations 
of class 3 and one of class 2. This means that our original coordinates were 
$\delta$-regular and we can use our first coordinate system in order 
to obtain intrinsic results.

In order to see whether the symbol is involutive, we differentiate each
equation with respect to $x^{1},x^{2},x^{3}$ from which we can produce 
$\mathcal{M}_{2}$, where its sparse coefficient matrix is given and explained 
in detail in \cite{Ger1}. We can determine the rank of this sparse matrix 
easily by hand (or using Maple) and obtain $r(\mathcal{M}_{2})=18$ as shown in
\cite{Ger1}. We now compute the total number of multiplicative variables 
$$\sum^{3}_{k=1}k\cdot\beta^{(k)}_{1}=17 \neq r(\mathcal{M}_{2})=18\: ,$$ 
which means that even the necessary condition for the symbol to be involutive 
is not fulfilled and the system cannot be in involution. 

Because $\mathcal{M}_{1}$ is not involutive, we prolong $\mathcal{R}_{1}$ to
$\mathcal{R}_{2}$ which consists of 24 equations 18 of which are the partial
derivatives of the 6 $f^{(R)}_{abcd}$. We write the prolonged symbol 
$\mathcal{M}_{2}$ down which is based on the 
following ranking. We order the $S_{abcde}$ such that $S_{abc33} \succ 
S_{abc23} \succ S_{abc22} \succ S_{abc13} \succ S_{abc12} \succ S_{abc11}$ and
then amongst each set $S_{233ij} \succ S_{232ij} \succ S_{132ij} \succ 
S_{123ij} \succ S_{133ij} \succ S_{131ij} \succ S_{122ij} \succ S_{121ij}$. 
We find that the symbol in orthonomic form is given by (with multiplicative 
variables indicated in each equation)
\[ \baar{llr}
\fbox{1} \qquad V_{12112} & = V_{12211} & \qquad x^{1} \: \bullet \: 
\bullet \nn\\
\fbox{2} \qquad V_{12122} & = V_{12212} & \qquad x^{1} \: x^{2} \: 
\bullet \nn\\
\fbox{3} \qquad V_{12123} & = 2V_{12312}-V_{13212}+V_{23211} & \qquad
 x^{1} \: x^{2} \: \bullet \nn\\
 & & \nn\\
\fbox{4} \qquad V_{13113} & = V_{13311} & \qquad x^{1} \: \bullet \: 
\bullet \nn\\
\fbox{5} \qquad V_{13123} & = V_{13312} & \qquad x^{1} \: x^{2} \: 
\bullet \nn\\
\fbox{6} \qquad V_{13133} & = V_{13313} & \qquad x^{1} \: x^{2} \: x^{3} \nn\\
 & & \nn\\
\fbox{7} \qquad V_{23213} & = V_{23312} & \qquad x^{1} \: \bullet \: 
\bullet \nn\\
\fbox{8} \qquad V_{23223} & = V_{23322} & \qquad x^{1} \: x^{2} \: 
\bullet \nn\\
\fbox{9} \qquad V_{23233} & = V_{23323} & \qquad x^{1} \: x^{2} \: x^{3} \nn\\
 & & \nn\\
\fbox{10} \qquad V_{12113} & = V_{12311}-V_{13112}+V_{13211} & \qquad x^{1} 
\: \bullet \: \bullet \nn\\
\fbox{11} \qquad V_{13122} & = 2V_{13212}-V_{12312}-V_{23211} & \qquad
 x^{1} \: x^{2} \: \bullet \nn\\
\fbox{12} \qquad V_{12133} & = \frac{3}{2}V_{12313}-\frac{1}{2}
    V_{13312}+\frac{1}{2}V_{23311} & \qquad x^{1} \: x^{2} \: x^{3}
\eaar \]
\[ \baar{llr}
\fbox{13} \qquad V_{12213} & = 2V_{12312}-V_{13212}+V_{23211} & \qquad
 x^{1} \: \bullet \: \bullet \nn\\
\fbox{14} \qquad V_{12223} & = 2V_{12322}-V_{13222}+V_{23212} & \qquad
 x^{1} \: x^{2} \: \bullet \nn\\
\fbox{15} \qquad V_{12233} & = \frac{3}{2}V_{12323}+\frac{1}{2}V_{23312}
-\frac{1}{2}V_{13322} & \qquad x^{1} \: x^{2} \: x^{3} \nn\\
 & & \nn\\
\fbox{16} \qquad V_{13213} & = \frac{1}{2}(V_{13312}+V_{12313}+V_{23311}) 
& \qquad x^{1} \: \bullet \: \bullet \nn\\
\fbox{17} \qquad V_{13223} & = \frac{1}{2}(V_{13322}+V_{12323}+V_{23312})
& \qquad x^{1} \: x^{2} \: \bullet \nn\\
\fbox{18} \qquad V_{13233} & = \frac{1}{2}(V_{13323}+V_{12333}+V_{23313}) 
& \qquad x^{1} \: x^{2} \: x^{3} \label{5Rie3S2} .
\eaar \]
This system produces 6 equations of class 1, 7 of class 2 and 5 of class 3
leading to $\beta^{(1)}_{2}=6,\beta^{(2)}_{2}=7,\beta^{(3)}_{2}=5$.
The total number of multiplicative variables equals 35 and if 
$\mathcal{M}_{2}$ were involutive, we would have to obtain 
$r(\mathcal{M}_{3})=35$. 
Differentiating $\mathcal{R}_{2}$ with respect to all
3 spacetime coordinates, we obtain 54 equations relevant for 
the symbol $\mathcal{M}_{3}$. But all those equations obtained from
differentiation with respect to non-multiplicative variables are
superfluous as each of them can be obtained by differentiating with
respect to multiplicative variables except for $\pr_{3}f^{(R)}_{1212}$. 
Therefore, $r(\mathcal{M}_{3}) \leq 36$, when we use formal differentiation of
the above symbol equations directly because $\pr_{3}(eqn. 3)$ corresponding to
$\pr_{3} f^{(R)}_{1212}$ is a linear combination of some other equations in 
$\mathcal{M}_{3}$. We obtain $r(\mathcal{M}_{3})=35$, which means that 
$\mathcal{M}_{2}$ is involutive.

We obtain that the symbol equation $\pr_{3}(eqn. 3)$ can also be created as a
linear combination of
\be
\pr_{3}(eqn. 3) = 
2 \pr_{2}(eqn. 16)+2 \pr_{3}(eqn. 11)-2 \pr_{3}(eqn. 13)- \pr_{2}(eqn. 5)
- \pr_{1}(eqn. 7) \label{inti}
\ee
using formal differentiation so that $r(\mathcal{M}_{3})=\sum^{3}_{k=1} k 
\cdot \beta^{(k)}_{2} = 35$. However, it turns out that when (\ref{inti}) is 
rewritten in terms of the full equations from $\mathcal{R}_{3}$ as
\be
I = f^{(R)}_{1212,33}+f^{(R)}_{1313,22}+f^{(R)}_{2323,11}-2f^{(R)}_{1323,12}
-2f^{(R)}_{1213,23}+2f^{(R)}_{1223,13} \label{finti} \: ,
\ee
it is not a trivial identity any longer and cannot be obtained by means of 
any linear combination of the $f^{(R)}_{abcd}$ and their derivatives 
$f^{(R)}_{abcd,e}$ so that $r(\mathcal{R}_{3})=6+18+36=60$. Because our system
is linear, it is instructive to compute
\[ \baar{ll}
\mbox{dim} (\mathcal{R}_{1})=18 \: , & \quad \mbox{dim} (\mathcal{M}_{1})=18 
\: , \nn\\
\mbox{dim} (\mathcal{R}_{2})=56 \: , & \quad \mbox{dim} (\mathcal{M}_{2})=30 
\: , \nn\\
\mbox{dim} (\mathcal{R}_{3})=100 \: , & \quad \mbox{dim} (\mathcal{M}_{3})=45
\: , \eaar \]
which leads us to the important consequence that
$$ \mbox{dim} (\mathcal{R}^{(1)}_{2})= \mbox{dim} (\mathcal{R}_{3})
- \mbox{dim} (\mathcal{M}_{3}) = 100 - 45 =55 = \mbox{dim} (\mathcal{R}_{2})
-1 \: .$$
But this means that {\bf one integrability condition of the form (\ref{finti}) 
occurs}. Our new system of equations has to be given by
\bear
0 & = & f^{(R)}_{abcd} \nn\\
0 & = & f^{(R)}_{abcd,e} \nn\\
0 & = & I \: . \label{proli}
\eear
We find that its symbol is simply given by the previous $\mathcal{M}_{2}$ 
together with
\be
V_{12323} = \frac{1}{\G^{3}_{12}-\G^{3}_{13}}[V_{23312}(\G^{3}_{13}
-\G^{3}_{12})+2\G^{3}_{33}V_{12312}-V_{13322}(2\G^{1}_{11}+\G^{3}_{12}
+\G^{3}_{13})] \label{derri}
\ee
which adopts 3 multiplicative variables. Therefore: 
$$\beta^{(1)}_{2,(1)}=6,\beta^{(2)}_{2,(1)}=7, \beta^{(3)}_{2,(1)}=6 \:\: 
\mbox{so that} \:\: \alpha^{(1)}_{2,(1)}=18,\alpha^{(2)}_{2,(1)}=9, 
\alpha^{(3)}_{2,(1)}=2\: .$$ We see that 
$r(\mathcal{M}^{(1)}_{3})=r(\mathcal{M}_{3})+3=38$ because the 3 derivatives
resulting from (\ref{derri}) are linearly independent and because
$$r(\mathcal{R}^{(1)}_{3})=r(\mathcal{R}_{1})+r(\mathcal{R}^{(1)}_{2})+35+3
=63$$ which leads to $$\mbox{dim} (\mathcal{R}^{(2)}_{2})=\mbox{dim}
(\mathcal{R}^{(1)}_{3})-\mbox{dim}(\mathcal{M}^{(1)}_{3})=97-42=55
=\mbox{dim}(\mathcal{R}^{(1)}_{2}) \: .$$
Because $r(\mathcal{M}^{(1)}_{3})=\sum^{3}_{k=1} k \cdot \beta^{(k)}_{2,(1)}
=6+14+18=38 \: ,$ $\mathcal{M}^{(1)}_{2}$ is involutive and because 
$\mbox{dim}(\mathcal{R}^{(2)}_{2})=\mbox{dim}(\mathcal{R}^{(1)}_{2})=55 \: ,$
there are no integrability conditions that can occur and so the system 
$\mathcal{R}^{(1)}_{2}$ is in involution.
\subsection{A Covariant Formulation of the Integrability Condition for $n=3$}
Even though the above prolongation of the Riemann-Lanczos problem 
mathematically leads to a prolonged system in involution, this prolonged
system (\ref{proli}) does not respect general covariance. This is not 
satisfactory from a general relativity point of view and we are going to 
suggest a covariant version of the above prolongation. Instead of adding the 
partial derivatives to our equations, we look at the following new system of 
equations
\bear
0 & = & f^{(R)}_{abcd} \nn\\
0 & = & f^{(R)}_{abcd;e} \: . \label{CoViP}
\eear
The symbol of this modified system obviously coincides with that of the system
formed by $f^{(R)}_{abcd}, f^{(R)}_{abcd,e}$ and again, we obtain for the rank
of $\mathcal{M}_{3}$ that $r(\mathcal{M}_{3})=\sum^{3}_{k=1} k \cdot 
\beta^{k}_{2} = 35$ so that $\mathcal{M}_{2}$ is involutive. We now have to 
see whether the linear combination (\ref{inti}) also holds for the full 
covariant system here. Again, calculations show that this is not the case and 
we obtain a covariant version of the above integrability condition in solved 
form given by
\bear
I_{Cov} & = & f^{(R)}_{1212;33}+f^{(R)}_{1313;22}+f^{(R)}_{2323;11}
-f^{(R)}_{1323;12}-f^{(R)}_{1323;21}-f^{(R)}_{1213;23}-f^{(R)}_{1213;32} \nn\\
& & +f^{(R)}_{1223;13}+f^{(R)}_{1223;31} \: .
\eear
This condition can be rewritten again in solved form as
\be
I_{Cov}=B_{12123;3}+B_{13132;2}+B_{23231;1} \: \label{blubber},
\ee
where $B_{abcde}:=f^{(R)}_{ab[cd;e]}$ which amounts to the covariant 
derivatives of the Bianchi identities for the $R_{abcd}$ involved in 
(\ref{blubber}). We can rewrite $I_{Cov}$ \cite{Sch2} in a more concise form
using {\bf bivectors}, where we introduce the bivector-indices 
$\underline{A}:=12, \: \underline{B}:=31, \: \underline{C}:=23\: .$ The 
integrability condition $I_{Cov}$ can then be expressed as
\bear
I_{Cov} & = & f^{(R)}_{\underline{A}\underline{A};(33)}
+f^{(R)}_{\underline{B}\underline{B};(22)}+f^{(R)}_{\underline{C}
\underline{C};(11)}+2(f^{(R)}_{\underline{A}\underline{C};(13)}
+f^{(R)}_{\underline{B}\underline{C};(12)}+f^{(R)}_{\underline{A}
\underline{B};(23)}) \nn\\
& = & \sum_{x,y=1}^{3}\sum_{X,Y=1}^{3}f^{(R)}_{XY;(xy)} \: .
\eear
Our new covariantly prolonged system is then given by
\bear
0 & = & f^{(R)}_{abcd} \nn\\
0 & = & f^{(R)}_{abcd;e} \nn\\
0 & = & I_{Cov} \: . \label{blubbb}
\eear
It is $r(\mathcal{R}^{(1)}_{2})=6+19=25, r(\mathcal{R}^{(1)}_{3})=6+19+38=63$
as well as $r(\mathcal{M}^{(1)}_{2})$\\
$=19, r(\mathcal{M}^{(1)}_{3})=38$ so that
$\mathcal{M}^{(1)}_{3}$ coincides with the previous symbol for the 
prolongation involving partial derivatives. Again, it is $\mbox{dim}
(\mathcal{R}^{(2)}_{2})=\mbox{dim}(\mathcal{R}^{(1)}_{2})=55$. 
We find that the system (\ref{blubbb}) also consists of a system in involution 
and in addition to this also obeys general covariance. Therefore, we prefer
(\ref{blubbb}) as a prolongation to a second-order system in involution for 
the Riemann-Lanczos problem in 3 dimensions and we state the final proposition:
\bt{Proposition}\\
The Riemann-Lanczos problem in 3 dimensions is not in involution.
Its reduced characters $(s_{0}',s_{1}',s_{2}',s_{3}')$ are
(14,8,7,3). The differential gauge conditions do not change the 
non-involutivity and they modify the reduced characters 
$(s_{0}',s_{1}',s_{2}',s_{3}')$ only slightly to (17,8,7,0).

The 3-dimensional Riemann-Lanczos problem becomes involutive after 
just one prolongation which is obtained by adding either of the integrability 
conditions (\ref{finti}) or (\ref{blubber}) and now its Cartan characters 
$(\alpha^{(1)}_{2},\alpha^{(2)}_{2},\alpha^{(3)}_{2})$ for 
$\mathcal{R}^{(1)}_{2}$ are (18,9,2).
\et
\subsection{A Singular Solution for the 3-dimensional Reduced G\"{o}del 
Spacetime}
Lastly, we give a singular solution for the unprolonged 3-dimensional G\"{o}del
spacetime. In 3 dimensions, the reduced G\"{o}del spacetime can be 
characterised by the following line element
\be
d s^{2} =a^{2}(d t^{2}- d x^{2} + \frac{1}{2}e^{2x}d y^{2} 
+ 2e^{x}d t d y) \: .
\ee
Here, we use the notation $x^{1}:=t, x^{2}:=x, x^{3}:=y$ for convenience. 
The Riemann-Lanczos equations together with the 3 differential gauge 
conditions are given explicitely in \cite{Ger1} and are very similar to the 
equations in 4 dimensions. Here, we want to find a solution which only 
depends on $x$ leading to the {\it Ansatz}
\[ \baar{ll}
L_{txy} = C_{1}e^{x}\: , \quad & L_{tyx} = C_{2}e^{x}\: , \nn\\
 & \nn\\
L_{txt} = C_{3}\: , \quad & L_{xyy} = C_{7}e^{2x} \: , \nn\\
 & \nn\\
L_{txx} = C_{4}e^{-x}\: , \quad & L_{tyt} = C_{6} \: , \nn\\
 & \nn\\
L_{tyy} = C_{5}e^{x}\: , \quad & L_{xyx} = C_{8} \: ,\label{5God3ans}
\eaar \]
where $C_{1},...C_{8}$ are arbitrary constants. When inserting this into
the 6 independent Riemann Lanczos equations and the 3 differential gauge 
conditions, we obtain a solution which does not satisfy 
${L_{ty}}^{s}_{\: ;s}=0$ \cite{Ger1} but when for instance choosing 
$C_{5}=C_{6}=0$, we obtain the solution
\[ \baar{ll}
L_{txy} = -\frac{a^{2}}{8}e^{(x)^{2}}\: , \quad & L_{tyx} = 
\frac{a^{2}}{8}e^{x}\: , \nn\\
 & \nn\\
L_{txt} = -\frac{a^{2}}{8}\: , \quad & L_{xyy} =\frac{3a^{2}}{16}e^{2x}\: .
\eaar \]
Its singular solution manifold possesses 3-dimensional tangent spaces which 
are spanned by 
\bear
U & = & \frac{\pr}{\pr t}\: , \nn\\
V & = & \frac{\pr}{\pr x}  
+\frac{a^{2}}{8}e^{x}\frac{\pr}{\pr L_{txy}}
+\frac{a^{2}}{8}e^{x}\frac{\pr}{\pr L_{tyx}}
+\frac{3 a^{2}}{8}e^{2x}\frac{\pr}{\pr L_{xyy}} \nn\\
& & +\frac{1}{3}e^{x}(\frac{5a}{16}-\frac{1}{2}-e^{2x}\frac{3a}{16})
\frac{\pr}{\pr P_{txyx}}
-\frac{1}{3}e^{x}(\frac{5a}{16}-\frac{1}{2}-e^{2x}\frac{3a}{16})
\frac{\pr}{\pr P_{tyxx}} \nn\\
& & +\frac{3}{4a}e^{2x}\frac{\pr}{\pr P_{xyyx}}\: , \nn\\
Z & = & \frac{\pr}{\pr y} \: .
\eear
We conclude that for this solution manifold $s_{0}=s_{0}'=3$ while all other 
characters vanish.
\section*{Conclusion}

In 2 dimensions, the Riemann-Lanczos problem is very simple and we showed 
that it is always in involution. The general solution was given for both
possible choices of signatures, Lorentzian and Euclidean.

In 3 dimensions, a prolongation becomes necessary to make it a system in 
involution. An integrability condition based on the derivatives of the 
Bianchi identities occurs when we use Janet-Riquier theory and introduce the
second-order partial derivatives $S_{abcde}$ as new jet coordinates. 
A singular solution for the unprolonged problem for the reduced G\"{o}del 
spacetime is given.
\section*{Acknowledgements} Both authors wish to thank
D Hartley and W M Seiler for valuable discussions as well as Prof L S Xanthis.
We are grateful to Prof C Hoenselaers for pointing out references
\cite{Per1,Per2,Per3,Per4} to us. 

A Gerber would like to thank the Swiss National Science Foundation 
(SNSF) and the Dr Robert Thyll-D\"{u}rr Foundation.
\renewcommand{\theequation}{A.\arabic{equation}}
\setcounter{equation}{0}
\section*{Appendix A: Calculations for 2 Dimensions}
In this section we exhibit results and calculations which are necessary for 
the Riemann-Lanczos problem in 2 dimensions. First, $\delta_{3}, \cdots , 
\delta_{6}$ are given by
\bear
\delta_{3} & = & g^{11}(2\G^{1}_{11}+\G^{2}_{12})
                 +g^{12}(3\G^{1}_{12}+\G^{2}_{22})
                 +g^{22}\G^{1}_{22} \: , \\
\delta_{4} & = & g^{12}(\G^{1}_{11}+3\G^{2}_{12})
                 +g^{22}(\G^{1}_{12}+2\G^{2}_{22})
                 +g^{11}\G^{2}_{11}\: ,\nn\\
\delta_{5} & = & P_{1211}g^{11}_{\: \:,1}+P_{1221}g^{12}_{\: \:,1}
                 +P_{1212}g^{12}_{\: \:,1}+P_{1222}g^{22}_{\: \:,1}
                 -L_{121}\delta_{3,1}-L_{122}\delta_{4,1} \: ,\nn\\
\delta_{6} & = & P_{1211}g^{11}_{\: \:,2}+P_{1221}g^{12}_{\: \:,2}
                 +P_{1212}g^{12}_{\: \:,2}+P_{1222}g^{22}_{\: \:,2}
                 -L_{121}\delta_{3,2}-L_{122}\delta_{4,2} \nn \: .
\eear
The polar matrices when the differential gauge condition is included then 
look like
\[ \baar{|l|l|l|l|l|l|l|l|l|} \hline
equation & d x^{1} & d x^{2} & d L_{121} & d L_{122} & d P_{1211} 
& d P_{1212} & d P_{1221} & d P_{1222}\\ \hline
  & & & & & & & & \\
d f^{(R)}_{1212} & \alpha_{12121} & \alpha_{12122} & 2(\G^{1}_{12}
-\G^{2}_{22}) & -2(\G^{1}_{11}+\G^{2}_{12}) & 0 & -2 & 2 &
0\\
d {L_{12}}^{s}_{\: ;s} & \delta_{5} & \delta_{6} & -\delta_{3} &
-\delta_{4} & g^{11} & g^{12} & g^{12} & g^{22}\\
K_{121} & -P_{1211} & -P_{1212} & 1 & 0 & 0 & 0 & 0
& 0\\
K_{122} & -P_{1221} & -P_{1222} & 0 & 1 & 0 & 0 & 0
& 0\\
2(U\rfloor d K_{121}) & -U_{1211} & -U_{1212} & 0 & 0 &
U^{1} & U^{2} & 0 & 0\\
2(U\rfloor d K_{122}) & -U_{1221} & -U_{1222} & 0 & 0 &
0 & 0 & U^{1} & U^{2}\\ 
2(V\rfloor d K_{121}) & -V_{1211} & -V_{1212} & 0 & 0 &
V^{1} & V^{2} & 0 & 0\\
2(V\rfloor d K_{122}) & -V_{1221} & -V_{1222} & 0 & 0 &
0 & 0 & V^{1} & V^{2} \\ \hline
\eaar , \]
where $U$ forms a 1-dimensional integral element and $span\{U,V\}$ 
form a 2-dimensional one. The linear multipliers $A_{1}, \cdots , A_{6}$ 
and $B_{1}, \cdots , B_{6}$ are given by
\[ \baar{ll}
A_{1} = \frac{1}{2}A_{2}(\frac{U^{1}}{U^{2}}g^{22}-g^{12}), &
A_{2} = \frac{V^{2}-\frac{U^{2}}{U^{1}}V^{1}}{\delta_{7}},\\
 & \\
A_{3} = A_{2}(\delta_{3}-(\G^{1}_{12}-\G^{2}_{22})
             (\frac{U^{1}}{U^{2}}g^{22}-g^{12})), &
A_{4} = A_{2}(\delta_{4}+(\G^{1}_{11}+\G^{2}_{12})
             (\frac{U^{1}}{U^{2}}g^{22}-g^{12})),\\
 & \\
A_{5} = \frac{1}{U^{1}}(V^{1}-g^{11}A_{2}), &
A_{6} = -\frac{1}{U^{2}}g^{22}A_{2},\\
 & \\
B_{1} = \frac{1}{2}B_{2}(g^{12}-\frac{U^{2}}{U^{1}}g^{11}), &
B_{2} = \frac{V^{1}-\frac{U^{1}}{U^{2}}V^{2}}{\delta_{8}},\\ 
 & \\
B_{3} = B_{2}(\delta_{3}-(\G^{1}_{12}-\G^{2}_{22})
             (g^{12}-\frac{U^{2}}{U^{1}}g^{11})), &
B_{4} = B_{2}(\delta_{4}+(\G^{1}_{11}+\G^{2}_{12})
             (g^{12}-\frac{U^{2}}{U^{1}}g^{11})),\\
 & \\
B_{5} = -\frac{1}{U^{1}}g^{11}B_{2}, &
B_{6} = \frac{1}{U^{2}}(V^{2}-g^{22}B_{2}),\\
\eaar \]
where $$
\delta_{7}=\delta_{8}=
2g^{12}-\frac{U^{2}}{U^{1}}g^{11}-\frac{U^{1}}{U^{2}}g^{22} \: .$$
\renewcommand{\theequation}{B.\arabic{equation}}
\setcounter{equation}{0}
\section*{Appendix B: Calculations for 3 Dimensions}
The quantity $\alpha_{abcde}$ for the Riemann-Lanczos problem in 3 dimensions
is defined as
\bear
\alpha_{abcde} & = & \G^{n}_{ad,e}(L_{nbc}+L_{ncb})+\G^{n}_{bc,e}(L_{nad}
+L_{nda})-\G^{n}_{ac,e}(L_{nbd}+L_{ndb}) \nn\\
& & -\G^{n}_{bd,e}(L_{nac}+L_{nca})\: .
\eear
We obtain 18 different ${\tilde{\alpha}}_{abcde}$ for the 3-dimensional
Riemann-Lanczos equations when using the full polar matrix,
which are given as follows \cite{Ger1}\\
(for $e=1,2,3$): 
\bear
{\tilde{\alpha}}_{1212e} & = & R_{1212,e}
+\alpha_{1212e}+ 2(\G^{1}_{12}+\G^{2}_{22})P_{121e}-2(\G^{1}_{11}
+\G^{2}_{12})P_{122e}\nn\\
 & & +2\G^{3}_{22}P_{131e}+2\G^{3}_{11}P_{232e}
+2\G^{3}_{12}P_{123e}-4\G^{3}_{12}P_{132e}\nn\\
{\tilde{\alpha}}_{1313e} & = & R_{1313,e}
+\alpha_{1313e}+ 2\G^{2}_{33}P_{121e}+2(\G^{1}_{13}+\G^{3}_{33})
P_{131e}-2(\G^{1}_{11}+\G^{3}_{13})P_{133e}\nn\\
 & & -2\G^{2}_{11}P_{233e}-4\G^{2}_{13}P_{123e}+2\G^{2}_{13}P_{132e}\nn\\
{\tilde{\alpha}}_{2323e} & = & R_{2323,e}
+\alpha_{2323e}-2\G^{1}_{33}P_{122e}-2\G^{1}_{22}P_{133e}+2(\G^{2}_{23}
+\G^{3}_{33})P_{232e}\nn\\
 & & -2(\G^{2}_{22}+\G^{3}_{23})P_{233e}+2\G^{1}_{23}
P_{123e}+2\G^{1}_{23}P_{132e}\nn\\
{\tilde{\alpha}}_{1213e} & = & R_{1213,e}
+\alpha_{1213e}+ (\G^{1}_{13}+2\G^{2}_{23})P_{121e}-\G^{2}_{13}P_{122e}
+(\G^{1}_{12}+2\G^{3}_{23})P_{131e} \nn\\
& & -\G^{3}_{12}P_{133e}-\G^{2}_{11}P_{232e}-(\G^{1}_{11}+2\G^{2}_{12}
-\G^{3}_{13})P_{123e}-(\G^{1}_{11}+2\G^{3}_{13}-\G^{2}_{12})P_{132e}\nn\\
{\tilde{\alpha}}_{1223e} & = & R_{1223,e}
+\alpha_{1223e} -\G^{1}_{23}P_{121e}+(2\G^{1}_{13}+\G^{2}_{23})P_{122e}
+\G^{1}_{22}P_{131e}-\G^{3}_{22}P_{133e} \nn\\
& & -(\G^{2}_{12}+2\G^{3}_{13})P_{232e}+\G^{3}_{12}P_{233e}
-(\G^{1}_{12}+2\G^{2}_{22}+\G^{3}_{23})P_{123e} \nn\\
& & +(2\G^{3}_{23}+2\G^{2}_{22}-\G^{1}_{12})P_{132e}\nn\\
{\tilde{\alpha}}_{1323e} & = & R_{1323,e}
+\alpha_{1323e}-\G^{1}_{33}P_{121e}+\G^{2}_{33}P_{122e}+\G^{1}_{23}P_{131e}
-(2\G^{1}_{12}+\G^{3}_{23})P_{133e} \nn\\
& & +\G^{2}_{13}P_{232e}-(\G^{3}_{13}+2\G^{2}_{12})P_{233e}
-(\G^{1}_{12}+2\G^{2}_{23}-\G^{3}_{33})P_{123e} \nn\\
& & +(\G^{1}_{13}+2\G^{3}_{33}+\G^{2}_{23})P_{132e}.
\eear
The polar matrix of $H(E^{3})_{x}$ for the 3-dimensional Riemann-Lanczos 
problem can be written as
\[ \left( \baar{lll}
M_{01} & M_{02} & M_{03} \nn\\
0 & M_{1} & 0 
\eaar \right) \: . \]
$M_{01}$ then is (for $d x^{1},d x^{2},d x^{3}$) the 6x3-matrix
\[ M_{01}:= \left( \baar{ccc}
R_{1212,1} + \alpha_{12121} \quad & R_{1212,2}+ \alpha_{12122} \quad & 
R_{1212,3}+\alpha_{12123}\\
R_{1313,1}+\alpha_{13131} \quad & R_{1313,2}+\alpha_{13132} \quad & 
R_{1313,3}+\alpha_{13133}\\
R_{2323,1}+\alpha_{23231} \quad & R_{2323,2}+\alpha_{23232} \quad & 
R_{2323,3}+\alpha_{23233}\\
R_{1213,1}+\alpha_{12131} \quad & R_{1213,2}+\alpha_{12132} \quad & 
R_{1213,3}+\alpha_{12133}\\
R_{1223,1}+\alpha_{12231} \quad & R_{1223,2}+\alpha_{12232} \quad & 
R_{1223,3}+\alpha_{12233}\\
R_{1323,1}+\alpha_{13231} \quad & R_{1323,2}+\alpha_{13232} \quad & 
R_{1323,3}+\alpha_{13233}
\eaar \right) \]
followed by the 8 columns (for the $d L_{abc}$) forming the 6x8-matrix 
$M_{02}$ which is
{\tiny \[ \left( \baar{cccccccc}
2(\G^{1}_{12}+ \G^{2}_{22}) &
-2(\G^{1}_{11}+\G^{2}_{12}) & 2 \G^{3}_{22} & 0 & 
2\G^{3}_{11} & 0 & 2 \G^{3}_{12} & -4\G^{3}_{12}\\
2\G^{2}_{33} & 0 & 
2(\G^{1}_{13}+\G^{3}_{33}) & -2(\G^{1}_{11}+\G^{3}_{13}) & 0 & 
-2\G^{2}_{11} & -4\G^{2}_{13} & 2\G^{2}_{13}\\
0 & -2\G^{1}_{33} & 0 & -2\G^{1}_{22} &
2(\G^{2}_{23}+\G^{3}_{33}) & -2 (\G^{2}_{22}
+\G^{3}_{23}) & 2\G^{1}_{23} & 2 \G^{1}_{23}\\
\G^{1}_{13}+ 2\G^{2}_{23} & -\G^{2}_{13} &
\G^{1}_{12}+2\G^{3}_{23} & -\G^{3}_{12} & 
-\G^{2}_{11} & \G^{3}_{11} & -(\G^{1}_{11}+ 2\G^{2}_{12} 
-\G^{3}_{13}) & -(\G^{1}_{11}+2\G^{3}_{13}-\G^{2}_{12})\\
-\G^{1}_{23} & 2\G^{1}_{13} +\G^{2}_{23} &
\G^{1}_{22} & -\G^{3}_{22} & -(\G^{2}_{12}+2
\G^{3}_{13}) & \G^{3}_{12} & -(\G^{1}_{12}+ 2\G^{2}_{22} 
+\G^{3}_{23}) & 2\G^{3}_{23}+\G^{2}_{22}-\G^{1}_{12}\\
-\G^{1}_{33} & \G^{2}_{33} & \G^{1}_{23} &
-(2\G^{1}_{12} +\G^{3}_{23}) & \G^{2}_{13} & 
-(\G^{3}_{13}+2\G^{2}_{12}) & 
-(-\G^{1}_{13}+ 2\G^{2}_{23} -\G^{3}_{33}) & 
\G^{1}_{13}+2\G^{3}_{33}+\G^{2}_{23}
\eaar \right) \] }
then followed by the 24 columns (for the $d P_{abcd}$) producing
the 6x24-matrix
\[ M_{03}:= \left( \baar{cccccccccccccccccccccccc}
0 & -2 & 0 & 2 & 0 & 0 & 0 & 0 & 0 & 0 & 0 & 0 & 0 & 0 & 0 & 0 & 0 & 0
& 0 & 0 & 0 & 0 & 0 & 0\\
0 & 0 & 0 & 0 & 0 & 0 & 0 & 0 & -2 & 2 & 0 & 0 & 0 & 0 & 0 & 0 & 0 & 0
& 0 & 0 & 0 & 0 & 0 & 0\\
0 & 0 & 0 & 0 & 0 & 0 & 0 & 0 & 0 & 0 & 0 & 0 & 0 & 0 & -2 & 0 & 2 & 0
& 0 & 0 & 0 & 0 & 0 & 0\\
0 & 0 & -1 & 0 & 0 & 0 & 0 & -1 & 0 & 0 & 0 & 0 & 0 & 0 & 0 & 0 & 0 & 0
& 1 & 0 & 0 & 1 & 0 & 0\\
0 & 0 & 0 & 0 & 0 & -1 & 0 & 0 & 0 & 0 & 0 & 0 & 1 & 0 & 0 & 0 & 0 & 0
& 0 & 2 & 0 & 0 & -1 & 0\\
0 & 0 & 0 & 0 & 0 & 0 & 0 & 0 & 0 & 0 & 1 & 0 & 0 & 0 & 0 & 1 & 0 & 0
& 0 & 0 & 1 & 0 & 0 & -2
\eaar \right) \: . \]  
$M_{1}$ consists of a 32x35-matrix which can be split into\\
\[ M_{1} := \left( \baar{ccc}
M(P) & 1 & 0\\
M(U) & 0 & N(U)\\
M(V) & 0 & N(V)\\
M(Z) & 0 & N(Z)
\eaar \right) \: . \]
M(P) is the 3x8-matrix\\
\[ M(P):= \left( \baar{ccc}
-P_{1211} & -P_{1212} & -P_{1213}\\
-P_{1221} & -P_{1222} & -P_{1223}\\
-P_{1311} & -P_{1312} & -P_{1313}\\ 
-P_{1331} & -P_{1332} & -P_{1333}\\ 
-P_{2321} & -P_{2322} & -P_{2323}\\
-P_{2331} & -P_{2332} & -P_{2333}\\
-P_{1231} & -P_{1232} & -P_{1233}\\ 
-P_{1321} & -P_{1322} & -P_{1323}
\eaar \right) \: . \]
M(U) is the 3x8-matrix\\
\[ M(U):= \left( \baar{ccc}
-U_{1211} & -U_{1212} & -U_{1213}\\
-U_{1221} & -U_{1222} & -U_{1223}\\
-U_{1311} & -U_{1312} & -U_{1313}\\
-U_{1331} & -U_{1332} & -U_{1333}\\
-U_{2321} & -U_{2322} & -U_{2323}\\
-U_{2331} & -U_{2332} & -U_{2333}\\
-U_{1231} & -U_{1232} & -U_{1233}\\
-U_{1321} & -U_{1322} & -U_{1323}
\eaar \right) \: . \]
N(U) then is the 8x24-matrix\\
\small {\[ \left( \baar{cccccccccccccccccccccccc}
U^{1} & U^{2} & U^{3} & 0 & 0 & 0 & 0 & 0 & 0 & 0 & 0 & 0 & 0 & 0 & 0
& 0 & 0 & 0 & 0 & 0 & 0 & 0 & 0 & 0\\
0 & 0 & 0 & U^{1} & U^{2} & U^{3} & 0 & 0 & 0 & 0 & 0 & 0 & 0 & 0 & 0 
& 0 & 0 & 0 & 0 & 0 & 0 & 0 & 0 & 0\\
0 & 0 & 0 & 0 & 0 & 0 & U^{1} & U^{2} & U^{3} & 0 & 0 & 0 & 0 & 0 & 0 
& 0 & 2 & 0 & 0 & 0 & 0 & 0 & 0 & 0\\
0 & 0 & 0 & 0 & 0 & 0 & 0 & 0 & 0 & U^{1} & U^{2} & U^{3} & 0 & 0 & 0 
& 0 & 0 & 0 & 0 & 0 & 0 & 0 & 0 & 0\\
0 & 0 & 0 & 0 & 0 & 0 & 0 & 0 & 0 & 0 & 0 & 0 & U^{1} & U^{2} & U^{3}
& 0 & 0 & 0 & 0 & 0 & 0& 0 & 0 & 0\\
0 & 0 & 0 & 0 & 0 & 0 & 0 & 0 & 0 & 0 & 0 & 0 & 0 & 0 & 0 & U^{1} &
U^{2} & U^{3} & 0 & 0 & 0& 0 & 0 & 0\\
0 & 0 & 0 & 0 & 0 & 0 & 0 & 0 & 0 & 0 & 0 & 0 & 0 & 0 & 0 & 0 & 0 & 0
& U^{1} & U^{2} & U^{3} & 0 & 0 & 0\\
0 & 0 & 0 & 0 & 0 & 0 & 0 & 0 & 0 & 0 & 0 & 0 & 0 & 0 & 0 & 0 & 0 & 0
& 0 & 0 & 0 & U^{1} & U^{2} & U^{3}
\eaar \right) \: . \]}
M(V) and M(Z) are the same as M(U) only for the Vessiot vector fields $V$ 
and $Z$. In the same way N(V) and N(Z) are the same as N(U).

Next, we give the linear multipliers creating a linear combination for 
the reduced polar matrix.
\[ \baar{lll}
A_{1}=\frac{\delta_{1}}{2}B_{1}\: , & A_{2}=\frac{\delta^{2}_{2}}
{2\delta_{1}}B_{1}\: , &
A_{3}=-\frac{\delta^{2}_{4}V^{2}}{2\delta^{2}_{3}V^{1}}B_{1} \: ,\nn\\
 & & \nn\\ 
A_{4}=\delta_{2}B_{1}\: , &
A_{5}=\frac{V^{2}\delta_{4}}{V^{1}}B_{1} \: , & 
A_{6}=\frac{V^{2}\delta_{2}\delta_{4}}{V^{1}\delta_{1}}B_{1}\: ,\\
 & & \nn\\
B_{1}=B_{1}(arbitrary) \: , & B_{2}=\frac{V^{2}}{V^{1}}B_{1}\: , &
B_{3}=\frac{\delta_{2}}{\delta_{1}}B_{1} \: ,\nn\\
 & & \nn\\ 
B_{4}=\frac{V^{3}\delta_{2}}{V^{2}\delta{1}}B_{1} \: , & 
B_{5}=-\frac{V^{2}\delta_{4}}{V^{1}\delta_{3}}B_{1}\: , & 
B_{6}=-\frac{V^{3}\delta_{4}}{V^{2}\delta_{3}}B_{1}\: ,\nn\\
 & & \nn\\
B_{7}=\frac{V^{2}}{V^{1}}(\frac{\delta_{2}}{\delta_{1}}-2\delta_{4})B_{1}
\:,
 & B_{8}=-\frac{1}{\delta_{3}}(\delta_{2}+\delta_{4})B_{1} \: ,\nn\\
 & & \nn\\
C_{1}=-\frac{U^{1}}{V^{1}}B_{1}\: , &
 C_{2}=-\frac{U^{2}}{V^{1}}B_{1}\: , &
C_{3}=-\frac{U^{1}\delta_{2}}{V^{1}\delta_{1}}B_{1}\: ,\nn\\
 & & \nn\\
C_{4}=-\frac{U^{3}\delta_{2}}{V^{1}\delta_{1}}B_{1} \: , & 
C_{5}=\frac{U^{2}\delta_{4}}{V^{1}\delta_{3}}B_{1}\: , & 
C_{6}=\frac{U^{3}\delta_{4}}{V^{1}\delta_{3}}B_{1}\: ,\nn\\
 & & \nn\\
C_{7}=-\frac{1}{V^{1}}(\delta_{2}+U^{1}V^{2}\frac{\delta_{2}}{\delta_{1}}
-2\frac{V^{2}}{V^{1}}U^{1}\delta_{4})B_{1}, & 
C_{8}=(\frac{\delta_{4}}{V^{1}}+\frac{U^{2}\delta_{4}}{V^{2}\delta_{3}}
+\frac{U^{2}\delta_{2}}{V^{2}\delta_{3}})B_{1}, &
\eaar \]
where here
\[ \baar{lll}
\delta_{1}=U^{2}-V^{2}\frac{U^{1}}{V^{1}}\: , \quad &
\delta_{2}=U^{3}-V^{3}\frac{U^{1}}{V^{1}}\: , \quad &
\delta_{3}=U^{1}-V^{1}\frac{U^{2}}{V^{2}}\: , \quad \nn\\
 & & \nn\\
\delta_{4}=U^{3}-V^{3}\frac{U^{2}}{V^{2}}\: , \quad &
\delta_{5}=U^{1}-V^{1}\frac{U^{3}}{V^{3}}\: , \quad &
\delta_{6}=U^{2}-V^{2}\frac{U^{3}}{V^{3}}\: , \quad 
\eaar \]
and
$$ \delta_{3}=-\frac{V^{1}}{V^{2}}\delta_{1}\: , \quad 
   \delta_{5}=-\frac{V^{1}}{V^{3}}\delta_{2}\: , \quad
   \delta_{6}=-\frac{V^{2}}{V^{3}}\delta_{4} \: . $$
For the special choice $B_{1}=U^{1}V^{1}V^{2}$, 
we obtain a similar kind of identity as the one mentioned in (23) in 
\cite{Bam1} for the 4-dimensional case but not with exactly the same 
multipliers because we incorporated the cyclic conditions (\ref{1cyclic}) as 
well.
\renewcommand{\theequation}{C.\arabic{equation}}
\setcounter{equation}{0}
\section*{Appendix C: A Coordinate Transformation for the 3-dimensional Case}
The $B_{i}$ in (\ref{5Rie3del}) are the parts which do not involve any term 
of the form $V_{abc3}$ and are given by
\bear
B_{1} & = & a_{21}V_{1211}+a_{22}V_{1212}-a_{11}V_{1331}-a_{12}V_{1222} \nn\\
B_{2} & = & a_{31}V_{1311}+a_{32}V_{2322}-a_{11}V_{1331}-a_{12}V_{2332} \nn\\
B_{3} & = & a_{31}V_{2321}+a_{32}V_{2322}-a_{21}V_{2331}-a_{22}V_{2332} \nn\\
B_{4} & = & a_{31}V_{1211}+a_{32}V_{1212}-a_{11}V_{1231}-a_{22}V_{1232}
           +a_{21}V_{1311}+a_{22}V_{1312} \nn\\
& & -a_{11}V_{1321}-a_{12}V_{1232} \nn\\
B_{5} & = & a_{31}V_{1221}+a_{32}V_{1222}-2a_{21}V_{1231}-2a_{22}V_{1232}
           +a_{21}V_{1321}+a_{22}V_{1322} \nn\\
& & -a_{11}V_{2321}-a_{12}V_{2322} \nn\\
B_{6} & = & 2a_{31}V_{1321}+2a_{32}V_{1322}-a_{21}V_{1331}-a_{22}V_{1332}
           -a_{31}V_{1231}-a_{32}V_{1232} \nn\\
& & -a_{11}V_{2331}-a_{12}V_{2332} \: .
\eear
The function $f(V_{abc1},V_{abc2})$ for $V_{2332}=f(V_{abc1},V_{abc2})$ is 
given by
\bear
V_{2332} & = &  \frac{a_{23}}{a_{23}a_{12}-a_{22}a_{13}}
[V_{1211}(\frac{a_{31}a_{33}}{a_{13}}-\frac{a_{21}{a_{33}}^{2}}{a_{13}a_{23}})
+V_{1212}(\frac{a_{33}a_{33}}{a_{13}}-\frac{a_{22}{a_{33}}^{2}}{a_{13}a_{23}})
\nn\\
& & +V_{1221}(\frac{a_{11}{a_{33}}^{2}}{a_{13}a_{23}}-\frac{a_{31}a_{33}}
{a_{23}})+V_{1222}(\frac{a_{12}{a_{33}}^{2}}{a_{13}a_{23}}
-\frac{a_{32}a_{33}}{a_{23}}) +V_{1311} \nn\\
& & (\frac{a_{21}a_{33}}{a_{13}}-\frac{a_{21}a_{33}}{a_{13}})
+V_{1312}(\frac{a_{22}a_{33}}{a_{13}}-\frac{a_{23}a_{33}}{a_{23}})
+V_{1331}(\frac{a_{11}a_{33}}{a_{13}}-a_{21}) \nn\\
& & +V_{1332}(a_{12}-a_{22})+V_{2321}(\frac{a_{11}a_{33}}{a_{23}}
-\frac{a_{13}a_{31}}{a_{23}})
+V_{2322}(\frac{a_{12}a_{33}}{a_{23}}-\frac{a_{13}a_{32}}{a_{23}}) \nn\\
& & +V_{2331}(\frac{a_{13}a_{21}}{a_{23}}-a_{11})
+V_{1231}(2\frac{a_{21}a_{33}}{a_{23}}-\frac{a_{11}a_{33}}{a_{23}}-a_{31})
+V_{1232} \nn\\
& & (2\frac{a_{22}a_{33}}{a_{23}}-\frac{a_{12}a_{33}}{a_{13}-a_{32}})
+V_{1321}(2a_{31}-\frac{a_{11}a_{33}}{a_{13}}-\frac{a_{33}a_{21}}{a_{23}})\nn\\
& & +V_{1322}(2 a_{32}-\frac{a_{12}a_{33}}{a_{13}}-\frac{a_{22}a_{33}}
{a_{23}}) \: .
\eear
\vspace{0.3cm}
\footnoterule
\vspace{0.2cm}\noindent
$^{1}$ All monomials which are not in the set of principal monomials or in its
extension by multiplicative variables correspond to parametric derivatives.\\
$^{2}$ Note that if a system of partial differential equations is translated 
into an exterior differential system in the appropriate way, then we have the 
correspondence\\
$s_{k}=\alpha^{(k)}_{q}$ as long as it is in involution. We 
refer the reader to \cite{Kura,5man} for further details concerning  
{\it involution}.
\bibliography{refs}
\end{document}